\RequirePackage{fix-cm}
\documentclass[twocolumn]{svjour3}          % twocolumn
\smartqed  % flush right qed marks, e.g. at end of proof

\usepackage{graphicx}
\usepackage{textcomp}
\usepackage{amsmath}
\usepackage{enumitem}
\usepackage{picins}
\usepackage{lineno,hyperref}

\usepackage[final]{changes}

\newcommand{\stkout}[1]{\ifmmode\text{\sout{\ensuremath{#1}}}\else\sout{#1}\fi}
\setdeletedmarkup{\stkout{#1}}

\usepackage{array}

\newcolumntype{L}[1]{>{\raggedright\let\newline\\\arraybackslash\hspace{0pt}}m{#1}}
\newcolumntype{C}[1]{>{\centering\let\newline\\\arraybackslash\hspace{0pt}}m{#1}}
\newcolumntype{R}[1]{>{\raggedleft\let\newline\\\arraybackslash\hspace{0pt}}m{#1}}

\begin{document}

\title{Fuzzy Inference Procedure for Intelligent and Automated Control of Refrigerant Charging\\ . \\The link to the formal publication is via\\
	{\small \url{https://doi.org/10.1007/s40815-018-0486-3}}%\thanks{Grants or other notes
}


\author{
}

\author{Issam W. Damaj \and Jean J. Saade \and Hala R. Al-Faisal \and Hassan B. Diab
}

\authorrunning{Short form of author list} % if too long for running head

\institute{Issam W. Damaj \at
              Department of Electrical and Computer Engineering, American University of Kuwait, Salmiya, Kuwait\\
              Tel.: +9651802040\\
              \email{idamaj@auk.edu.kw}           %  \\
           \and
           Jean J. Saade \at
             Department of Electrical and Computer Engineering, American University of Beirut, American University of Beirut, Beirut, Lebanon\\
           \and
           Hala R. Al-Faisal\at
             Department of Computer Engineering, Kuwait University, Khaldiya, Kuwait\\
           \and
           Hassan B. Diab\at
           Department of Electrical and Computer Engineering, American University of Beirut, Beirut, Lebanon
}

\date{Received: date / Accepted: date}

\maketitle

\begin{abstract}
\added{Fuzzy logic controllers are readily customizable in natural language terms, and can effectively deal with non-linearities and uncertainties in control systems.} This paper presents \deleted{an approach using fuzzy inference methodology to pave the way for the emergence of} an intelligent and automated \added{fuzzy} control procedure for the refrigerant charging of refrigerators. The elements that affect the experimental charging and the optimization of the performance of refrigerators are fuzzified and used in \added{an} \deleted{fuzzy} inference model. The objective is to represent the intelligent behavior of a human tester\added{,} and ultimately make \added{the developed} \deleted{this} model available for \added{the} use in an automated data acquisition, monitoring\added{,} and decision-making system\added{.} \added{The proposed system} is capable of determining the needed amount of refrigerant in the shortest possible time. \added{The system automates the refrigerant charging and performance testing of parallel units. The system is built using data acquisition systems from National Instruments and programmed under LabVIEW.} The \added{developed} fuzzy model\added{s,} and their testing results\added{,} are \added{evaluated according} \deleted{discussed and commented upon as} to their compatibility with the principles that govern the intelligent behavior of human experts when performing the refrigerant charging process. \added{In addition,} comparisons of the fuzzy models with classical inference models are \added{presented}\deleted{given}.\deleted{Furthermore, comments are provided on the ease by which the offered fuzzy model can be tuned to accommodate experts’ expectations in various refrigerant-charging scenarios to lead to reliable results.} The obtained results confirm that the proposed fuzzy controllers outperform traditional crisp controllers and provide major test time and energy savings. The paper includes \deleted{a} thorough \added{discussions}, analysis\added{,} and evaluation.

\end{abstract}

\section{Introduction}
\label{intro}
The manufacturing of refrigerators comprises thorough electromechanical safety and a variety of performance tests. Safety tests are performed per international standards and can include circuit, voltage, insulation, power, current leakage tests, etc. Well designed, reliable and validated performance tests are critical for quality production. Both safety and performance tests can be automatic or semi-automatic, and supported by a variety of equipment, data acquisition (DAQ) systems, and human operators \cite{GTP,AGRAMKOW}. 

Refrigerator performance tests vary in type and duration. Tests can be long-term with durations between 1 to 3 hours, or short-term with durations between 10 and 45 minutes. Further tests can be applied to samples of production over a duration between 3 and 12 hours. Tests in thermo-regulated rooms can last between 12 and 72 hours. Performance tests include temperature readings from the refrigerator main parts, such as, the compartments and evaporator, besides its ambient. Moreover, the tests include careful control of refrigerant charge quantities to attain the best possible performance and enable sufficient cooling and avoid various hazards \cite{Adler06}.

Due to the important effect of the amount of refrigerant charge on the performance of refrigerating units, the influence of the amount of refrigerant on the functioning of these units has been addressed in \cite{kuijpers1988influence,radermacher1996domestic}. Traditionally, the optimization of the amount of charge has been done experimentally and either manually or semi-automatically. However, refrigerant charging is proven to require time and labor and also consumes energy \cite{kuijpers1988influence,djd00}. In the literature, soft computing and classical methods to develop, computerize and automate the control and testing of refrigerators have been given a wide attention \cite{bandarra2016energy,yang2015self,kocyigit2015fault,you2012optimizing,li2011fuzzy,csahin2012comparative,rashid2010design,pang2016strategy}. The aims of the identified studies, however, differ from the objectives of this investigation.  

%GTP,AGRAMKOW,Adler06,kuijpers1988influence,radermacher1996domestic,djd00,

%Due to the important effect of the amount of refrigerant charge on the performance of refrigerating units, the determination of the required amount of refrigerant has been addressed in a variety of investigations . Traditionally, the optimization of the amount of charge has been done experimentally and either manually or semi-automatically. However, refrigerant charging is proven to require time and labor, and consumes energy . Indeed, analytical methods to develop, computerize and automate refrigerators testing, including the charging process, have been given wide attention 

The experimental setting of refrigerant charging optimization can become more appealing if the process is automated\added{,} while accounting for the reduction of the time, labor, and energy factors. Fuzzy logic and inference methodology can be employed to emulate the intelligent behavior of human experts in the control process. Fuzzy inference has been essentially developed for the design of humanistic or approximate reasoning systems \cite{zedeh1973outline,driankov1996introduction,zak2003systems}. Fuzzy control could fit well the human-performed and/or semi-automatic optimization of the refrigerant charging and testing processes. 

In this paper, we investigate the automation of the refrigerant charging and performance testing processes. The aims comprise saving time, labor, and energy while maintaining the processes adequacy. The investigation includes the development of a fuzzy inference, a classical control, and refrigerant charging and testing DAQ systems. The developed systems are implemented under LabVIEW and targets DAQ devices from National Instruments \cite{essick2013hands,NI}. The proposed system supports parallel analysis, testing, and charging of several refrigerators.  

This paper is organized so that Section~\ref{sec:2} presents related works. Section~\ref{sec:3} discusses the motivation and research objectives. In Section~\ref{sec:4}, background information is presented. In Section~\ref{sec:5}, the expert testing procedure is presented. Sections~\ref{sec:6} and~\ref{sec:7} present the developed classical and fuzzy solutions. Section~\ref{sec:8} presents the system architecture and implementation. \deleted{A} Thorough analysis and evaluation are presented in Section~\ref{sec:9}. Section~\ref{sec:10} concludes the paper and sets the ground for future work. \added{The Appendix include a list of acronyms and symbols, that are used in the paper, and their definitions.} 

\section{Related Work}
\label{sec:2}

Automated and fuzzy operation control, and performance testing, of refrigerators are widely addressed in the literature. The aim of investigations includes optimizing performance, simplifying control procedures, and/or mimicking the behavior of a human operator. Filho et al. in \cite{bandarra2016energy} developed an adaptive fuzzy controller for a 5-ton vapor compression system. The developed system achieved high coefficient of performance (\textit{COP}) values at relatively low rotation frequencies. Yang et al. in \cite{yang2015self} proposed a self-adjusting fuzzy controller to improve the performance of refrigerators and their power efficiency. The developed system supports self-adjusting the fuzzy rules in real-time. Kocyigit in \cite{kocyigit2015fault} successfully analyzed faulty conditions of a refrigeration system using fuzzy inference system and artificial neural networks. The developed system-fault and sensor-error analysis procedures are based on the thermodynamic properties and refrigeration cycles of the system. You et al. in \cite{you2012optimizing} investigated the optimization of the performance of refrigerators under large-scale load variations using fuzzy control. Significant energy saving is reported at an increased \textit{COP}. Saving energy and improving the \textit{COP} is also investigated by Li and Fei in \cite{li2011fuzzy}. The investigation included fuzzy control, for the capacity and superheat, to avoid the complexity of exact dynamic models and take the advantage of mimicking the human operator behavior with simplicity.

A classical step-wise optimization of charge amount is presented in \cite{pang2016strategy} for the Joule-Thomson cooler. The proposed optimization procedure successfully identified the best mixed-charge amounts. The lowest refrigeration temperature of -139 {\textcelsius} is obtained after 134 minutes of running time. The required amount of charge to achieve optimal performance, is commonly known to depend on the evaporator and condenser capacities, outer heat transfer surfaces, the capacity of the capillary tube, the ambient temperature, etc. \cite{radermacher1996domestic,dmitriyev1984determination}. 

Damaj et al. in \cite{djd00} investigate the refrigerant charging process using fuzzy control within a factory performance testing and quality control experimental setup. The developed fuzzy controller is basic and aims at mimicking the behavior of a human operator. Significant time savings in the testing procedure are reported while maintaining the adequacy of the \textit{COP} and the quality of the tested system. The study is limited to a single experiment setup and did not include an evaluation of the system impact on energy consumption or labor. In addition, the setup does not include an automated refrigerant charging system.

\section{Research Objectives}
\label{sec:3}
The purpose of the refrigerant charging process is to find the amount of refrigerant needed to attain the highest \textit{COP} at the best possible temperature. With no doubt, rapidly deciding on the optimal amount with enough assurance is of various advantages. Advantages include saving testing time, labor, and energy. Experimental charging is done through repeated addition of a specific amount of refrigerant\added{,} and monitoring the temperature change for each obtained amount as a function of time. The monitored temperature is usually taken as the average internal temperature of the refrigerator cabinet obtained by considering the low and high temperature sections as well as their capacities. Every time an amount of refrigerant is added, the minimum possible attainable temperature should be detected in as small time as possible.        

\begin{figure}
	%\centering
	\includegraphics[width=0.4\textwidth]{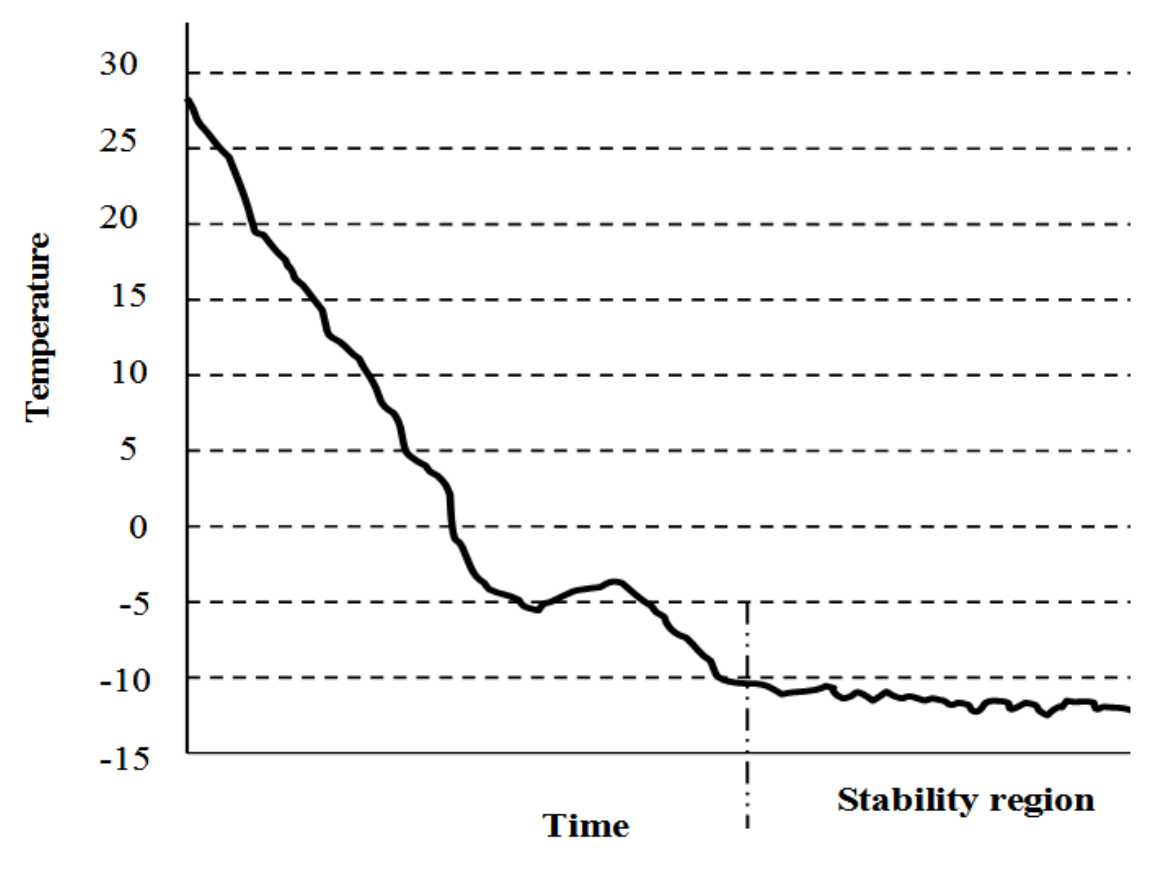}
	\caption{The change in temperature versus time.}
	\label{fig:1}       % Give a unique label
\end{figure}

In the targeted refrigeration systems, the regular trend of temperature change with time is decreasing in nature (See an approximation in Figure~\ref{fig:1}). Accordingly, if the period over which this change is observed, as it starts to become insignificant, is made shorter, then the risk of detecting a temperature larger than the optimum becomes higher. The way to deal with the stated problem is to place appropriate emphasis on the correlation between the length of the observation interval and the extent of the temperature change. The objective is to make the observation interval correspond closely and be fairly sufficient relative to the desired stable temperature range. The aim is to allow for the detection of the lowest or close to the lowest temperature\added{,} while not requiring too much time to finish the testing process.

Considering the problem statement and the concise description of its solution procedure, fuzziness arises naturally. For instance, fuzziness is involved in the terms: shorter, insignificant, much larger, fairly sufficient, close to the lowest, \added{and} too much time. These terms can be represented by fuzzy sets with suitable membership functions. In addition, fuzzy if-then inference rules can be structured to fit the problem description and provide appropriate solutions \cite{zedeh1973outline,driankov1996introduction,zak2003systems}. The fuzzy models that are to be devised would have to represent, therefore, how a human expert performs the refrigerant charging process. Inference rules, which can be employed in the fuzzy models, can be of the form shown in Equation~\ref{eqn:1}.

\begin{itemize}
	{\small
		\item  \textit{\textbf{If} temperature change is significant \textbf{and} observation time is insufficient, \textbf{then} testing continues.}
		\item  \textit{\textbf{If} temperature change is insignificant \textbf{and} observation time is fairly sufficient, \textbf{then} testing stops.}
	}
\end{itemize}
\begin{equation}
	\label{eqn:1}
\end{equation}

The purpose of this paper is to use fuzzy logic and inference methodologies to pave the way for the emergence of an intelligent and automatic control procedure for the process of refrigerant charging. A detailed description and more justification of the use of the fuzzy inference rules and the results they provide along with comments on these results are provided in Sections 4, 6 and 7. \added{Furthermore, since limited inference-based control methods for refrigerant charging of refrigerators have been reported in the literature, non-fuzzy inference models based on crisp sets or intervals are presented in Section 6 for comparison purposes. We should mention at this point that fuzzy and other soft computing techniques were introduced and used in the refrigeration applications \cite{bandarra2016energy,yang2015self,kocyigit2015fault,you2012optimizing,li2011fuzzy,csahin2012comparative,rashid2010design}. However, the aims of the reported work in the literature differ from the objectives of the subject addressed in this paper. In this paper, we go beyond the work presented in \cite{djd00} with focus on achieving better time savings, \added{higher performance}, investigating the impact on energy consumption and labor use, and providing a real-time system that is capable of testing parallel refrigerators inside or outside thermos-regulated rooms.} The motivations behind this paper are \added{summarized} as follows:

\begin{itemize}
	\item Automating the performance testing of refrigerators is a useful industrial application and a challenging process control procedure. \added{The automation targets a widely-used appliance and enables a broad impact}.  
	
	\item The refrigerant charging process can be time consuming, labor intensive, and consumes high energy in parallel-testing setups. \added{To that end, the opportunity for multiple improvements \added{are} present}.
	
	\item \added{The design of a tight control system requires access to the full state vector of the underlying system; in practice, this is not always feasible~\cite{PhysRevE.75.056203}. The presence of the human factor, in this problem, cannot be modeled or implemented using a classical controller with the usual mathematical formulations at a reduced complexity. Mimicking the behavior of a human operator, and exploiting the intrinsic fuzziness of the problem, promise significant reductions in the control system complexity $\textendash$ given the availability of modern rapid-prototyping tools and the exclusion of the human operator}.
	
	\item Automating the process of monitoring and refrigerant charging improves the reliability of the procedure with the elimination of the chance of human errors. Besides, it increases the safety of the process. 
\end{itemize}

The research objectives are summarized as follows:
\begin{itemize}
	\item Reduce the time of the refrigerant charging procedure and performance testing of refrigerators \added{,} while maintaining the quality of the product.
	
	\item Capture the intrinsic fuzziness of the process \added{to provide accurate control at reduced complexity}.
	
	\item Develop a fuzzy system that captures the decision-making process of a human operator.
	
	\item Develop a classical control model for the sake of results comparison and system evaluation.
	
	\item Develop a testing system using high-end DAQ equipment that support parallel testing of many refrigerators. The system includes sensors, interfaces, hardware, and application software graphical user interface (\textit{GUI}).
	
	\item Present a thorough analysis and evaluation of the proposed solution, \added{and investigate possible improvements in testing time, control performance, labor needs, and power consumption}.
\end{itemize}

\section{Background}
\label{sec:4}

\subsection{Fuzzy Sets and Fuzzy Logic}

The concept of a fuzzy set \cite{zadeh1965fuzzy} consists of a generalization of the concept of a classical set. An element in a space over which a fuzzy set is defined can have a partial grade of membership in a fuzzy set instead of being either a member or not a member as in a classical set. Hence, the notion of a characteristic function is generalized to become a membership function mapping the space over which the fuzzy set is defined to the interval [0, 1]. This makes the membership function represent a smooth rather than an abrupt transition from complete membership to non-membership. Here, a fuzzy set becomes an entity having no sharply defined boundary. 

\begin{figure}
	%\centering
	\includegraphics[width=0.4\textwidth]{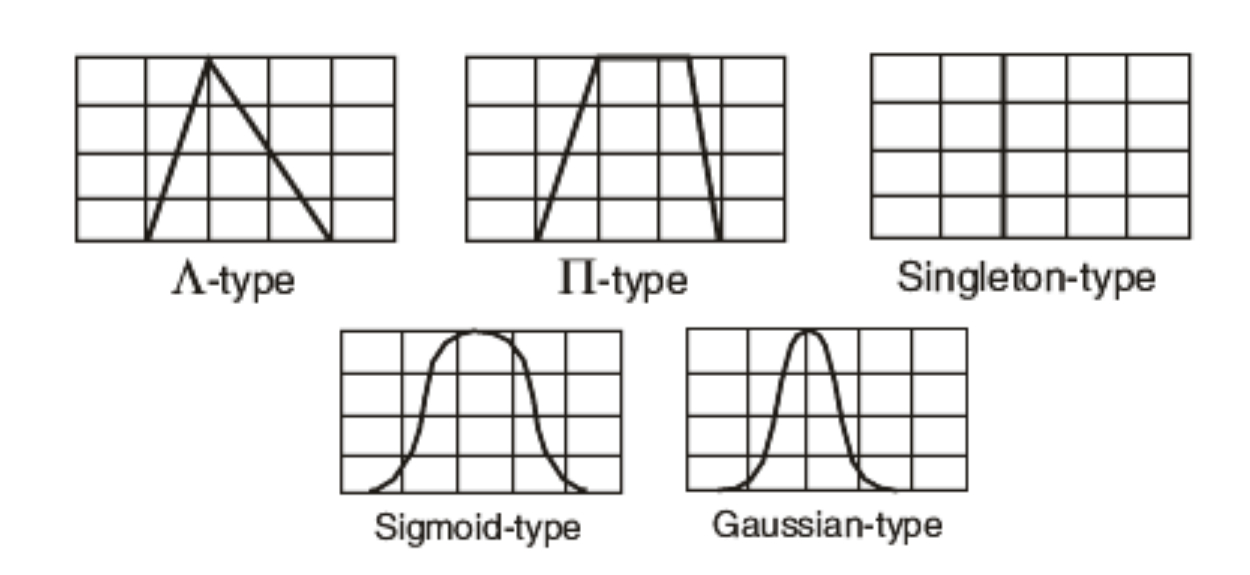}
	\caption{Different types of membership functions.}
	\label{fig:2}       % Give a unique label
\end{figure}

Fuzzy logic emerges from the concept of a fuzzy set to deal with fuzzy or vague linguistic propositions arising in natural languages. A simple fuzzy proposition, therefore, is not necessarily true or false. It can have a degree of truth in the interval [0, 1]. Some of the membership functions (\textit{MFs}) used to describe a fuzzy set are shown in Figure~\ref{fig:2} \cite{essick2013hands,zedeh1973outline,zadeh1975fuzzy,klir1995fuzzy,jiang2015improved,nguyen2012fuzzy}.
Simple fuzzy propositions can be combined by the logic operators \textit{AND} and \textit{OR}, for example, to yield composite propositions. The \textit{THEN} operator can also be used to construct conditional fuzzy propositions (fuzzy \textit{IF}-\textit{THEN} rules), which form the backbone of fuzzy inference systems (See Equation~\ref{eqn:1}) or fuzzy controllers.  

\subsection{Fuzzy Controllers}
\begin{figure}
	%\centering
	\includegraphics[width=0.4\textwidth]{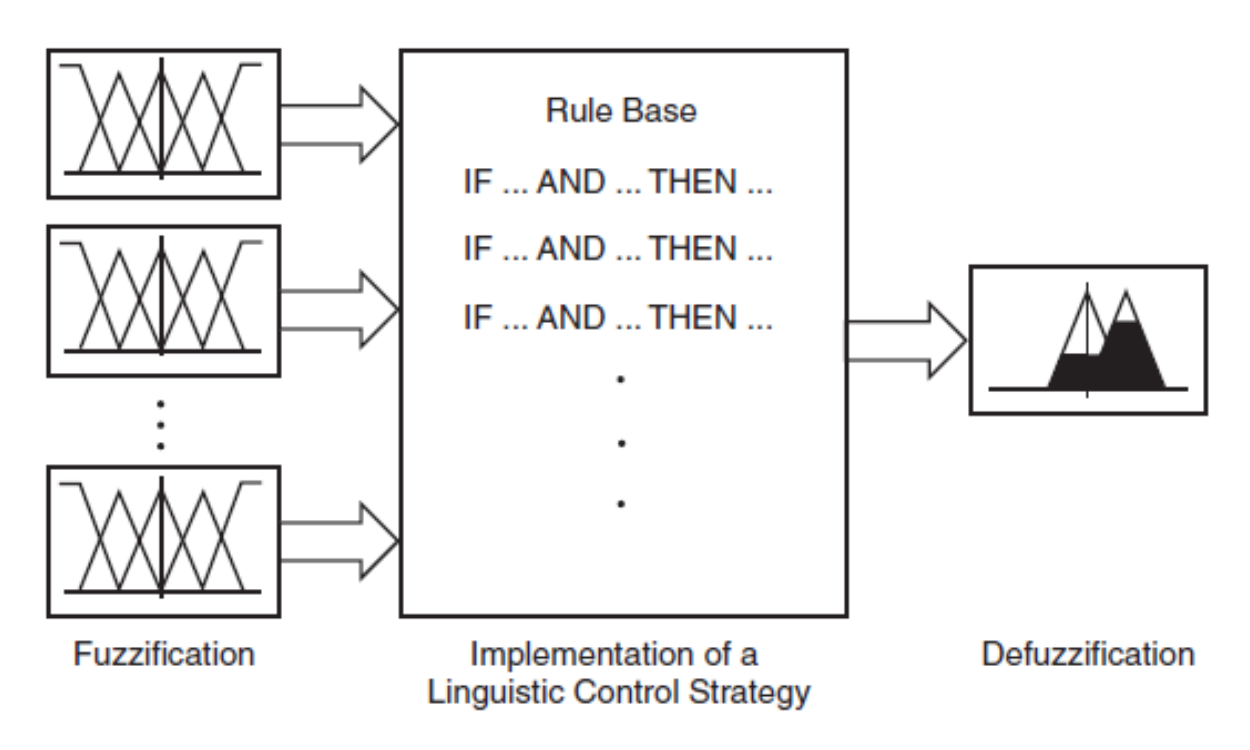}
	\caption{A fuzzy controller model.}
	\label{fig:3}       % Give a unique label
\end{figure}

A fuzzy controller is an input-output system formed by a fuzzifier, a defuzzifier and a collection of fuzzy inference rules (See Figure~\ref{fig:3}). Having fuzzy sets (overlapping ones) assigned over the input and output variables of the fuzzy controller, then the \textit{IF}-\textit{THEN} rules provide the necessary connection between these input and output fuzzy sets. For crisp input values, the membership grades in the input fuzzy sets are obtained by fuzzification. The fuzzified grades are then used in the rules setting and by applying the compositional rule of inference (\textit{CRI}) to obtain a corresponding fuzzy output, which needs to be defuzzified; i.e. converted into a crisp output. Indeed, various defuzzification methods are devised in the literature \cite{nguyen2012fuzzy,hellendoorn1993defuzzification,ccaugman2013intuitionistic,saade2000defuzzification}.

Let an N-rule, two-input, one-output fuzzy system be such that the $j^{th}$ rule, $1\leq j\leq N$, is expressed in Equation~\ref{eqn:2}.

\begin{equation}
	\label{eqn:2}
	R_j: \textbf{\textit{IF}} \: x \: is \: A_j \: \textbf{\textit{AND}} \: y \: is \: B_j, \: \textbf{\textit{THEN}} \: z \: is \: C_j\\
\end{equation}

$A_j$, $B_j$ and $C_j$ are fuzzy sets assigned over $x$, $y$ and $z$ which are, respectively, the input and output variables of the system. The \textit{CRI}-obtained fuzzy output, $C_0$, and corresponding to the crisp input pair $(x_0, \: y_0)$ is as in Equation~\ref{eqn:3}.

\begin{equation}
	\label{eqn:3}
	\mu _{C_0}(z)=\max _{1\leq j\leq N}[\mu _{Aj}(x_0)\wedge \mu _{Bj}(y_0)\wedge \mu _{Cj}(z)].
\end{equation}

The minimum operator ($\wedge$) is used to represent the $AND$ and $THEN$ operators. The $OR$ operator, that is usually considered between the rules, is represented by maximum ($max$). If crisp values, $C_j$’s, are assigned over the controller output variable, making the $THEN$ parts of the rules crisp, then a crisp output value corresponding to ($x_0, \: y_0)$ can be obtained directly from within the rules without passing a fuzzy output. The weighted average defuzzification formula, expressed in Equation~\ref{eqn:4}, is one way to obtain this crisp output \cite{zak2003systems,saade2000defuzzification}:

\begin{equation}
	\label{eqn:4}
	C_0=\frac{\sum ^{N}_{j=1}\mu _j\times c_j}{\sum ^{N}_{j=1}\mu_j}, where\ \mu _j=\mu _{Aj}(x_0)\wedge \mu _{Bj}(y_0)
\end{equation}

\subsection{Analysis and Design of Fuzzy Control Systems}
\added{The relationship between fuzzy logic and linear system theories motivates numerous investigations on the systematic analysis and design of fuzzy control systems. Accordingly, important investigations related to fuzzy system stability analysis have comprised component failures, reliable control, effects of time delays, chaotic modes, performance degradations, applications, etc.~\cite{7792142,7887738,7557043,7390222,6552850}.}

\section{Expert Testing Procedure}
\label{sec:5}

The usual refrigerant charging procedure is intrinsically fuzzy. Undoubtedly, the observation of temperature change, at various refrigerant amounts, and the time spent to attain this temperature, depend on the level of expertise of the hu-man tester. A conservative tester with less expertise would, for example, extend the observation time as much as possible and maybe beyond what is necessary to complete the process. In such a situation, the test could have been stopped at an “\textit{earlier time}” without having a significant effect on the reached temperature. Such an earlier and more adequate stop would mostly be done by a less conservative tester with a higher level of expertise. The experienced tester can correlate in a proficient manner between various ranges (or “\textit{approximate ranges}”) of temperature change and various corresponding observation time periods (or “\textit{approximate periods}”) to decide whether the test should continue or stop. 

In fact, the consideration of approximate ranges of temperature change and lengths of observation time periods can suitably be represented by fuzzy sets and logic. The fuzzy controller can be developed based-on how an expert interprets the curve in Figure~\ref{fig:1} to implement the above-mentioned correlation and translate it in the form of decision-making strategies. Now, the main question would therefore be: How should the change in temperature be concretely considered relative to the time that has elapsed to reach the desired point? In other words, how the stoppage or continuation of the testing process should be decided? 

It can be argued that the temperature change and observation time variables could be partitioned into ranges. Moreover, the stoppage or continuation of the testing procedure could be made to depend on each combination of two intervals, one for temperature change and the other for observation time. In fact, this constitutes the basis for a plausible classical solution whose main aspects are presented in Section~\ref{sec:6}. 

A more natural type of a solution and closer to how the testing procedure is usually performed by a human expert can be based on fuzzy logic and inference. Instead of partitioning the problem variables into intervals, the partitioning can be done into fuzzy intervals or equivalently fuzzy sets. Then, pair-wise combinations of these sets are to be structured and corresponding outputs, related to the stoppage (\textit{STOP}) or continuation (\textit{CONT}) of the testing, are to be assigned to give fuzzy inference rules. 

\section{Classical Solutions}
\label{sec:6}

In this section, and consistently with what has been presented in Section~\ref{sec:5}, two classical inference models are provided for the control of the refrigerant charging process. The first solution consists of partitioning the temperature change variable, denoted Dtemp, into two intervals: $0 \leq Dtemp< 0.5 \: \textcelsius$, and $Dtemp \geq 0.5 \: \textcelsius$. The observation time variable, denoted Time, is also partitioned into two intervals: $0 \leq Time < 30 \, minutes$ and $Time \geq 30 \, minutes$. The corresponding classical rules are presented in Equation~\ref{eqn:5}.\\

%\begin{itemize}
{\scriptsize
	\noindent  \textit{\textbf{If} $0 \leq Dtemp < 0.5$ \textcelsius \ \textbf{and} $Time \geq 30 \: min$, \textbf{then} $STOP$}
	
	\noindent \textit{\textbf{If} $0 \leq Dtemp < 0.5$ \textcelsius \ \textbf{and} $0 \leq Time < \: 30 \: min$, \textbf{then} $CONT$}
	
	\noindent \textit{\textbf{If} $Dtemp \geq 0.5$ \textcelsius \ \textbf{and} $Time \geq \: 30 \: min$, \textbf{then} $CONT$}
	
	\noindent \textit{\textbf{If} $Dtemp \geq 0.5$ \textcelsius \ \textbf{and} $0 \leq Time < 30 \: min$, \textbf{then} $CONT$}
	
}
%\end{itemize}
\begin{equation}
	\label{eqn:5}
\end{equation}

The implementation of the first classical solution considers the use of a timer that starts recording the time when $Dtemp$ reaches a value less than $0.5$. The scanning stops only when $Dtemp$ becomes less than $0.5$ {\textcelsius} for an observation time greater than or equal to 30 minutes. A charging situation is assumed here so that the noted observation period is considered as one that leaves no doubt about reaching the stability stage and thus recording the smallest temperature that can be achieved for a specific amount of refrigerant. 

\begin{figure}
	%\centering
	\includegraphics[width=0.4\textwidth]{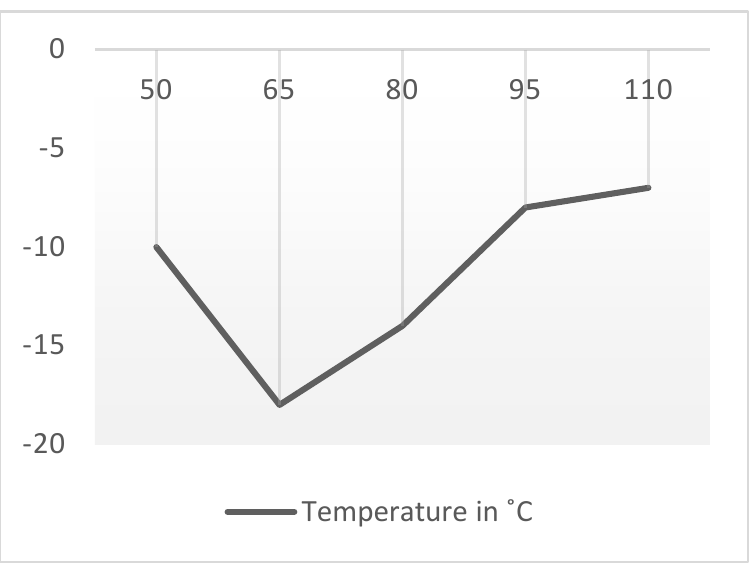}
	\caption{Refrigerant amount in grams vs. min. temperature.}
	\label{fig:4}       % Give a unique label
\end{figure}

More emphasis on the correspondence between the temperature change and observation time, as it relates to various refrigerant-charging situations, will be given in the sequel. When the scanning stops, another amount of refrigerant is added and the process is repeated. In Figure~\ref{fig:4}, a chart is drawn showing the refrigerant amount versus the minimum temperature. Thus, the amount of refrigerant needed to give the highest possible \textit{COP} can be found. But, this is not in as small a time as possible. Since, after reaching a value less than $0.5$ \textcelsius, $Dtemp$ could even become smaller requiring a stoppage time instant before 30 minutes. Such a possibility needs to be accounted for; this can be done as in a second classical controller.

An improvement to the solution in Equation~\ref{eqn:5} is to make it softer and hence closer to human expert’s testing (See Section~\ref{sec:5}). The improvement can be done by providing additional interval partitioning to the variables. For $Dtemp$, The intervals are: $[0, \: 0.25$ \textcelsius$)$, $[0.25$ \textcelsius$, \: 0.35$ \textcelsius$)$, $[0.35$ \textcelsius $, \: 0.5$ \textcelsius$)$ and $[0.5$ \textcelsius $, \: \infty)$. For Time, the intervals expressed in minutes are: $[0, \: 20)$, $[20, \: 25)$, $[25, \: 30)$ and $[30, \: \infty)$. Considering all the pair-wise combinations of these intervals and assigning an appropriate stop or continue testing for each combination, while respecting the principles of Section~\ref{sec:5}, then 16 rules are result (See Equation~\ref{eqn:6}).\\

%\begin{itemize*}[leftmargin=*]
{\scriptsize
	
	\noindent \textit{\textbf{If} $0 \leq Dtemp < 0.25$ \textcelsius \ \textbf{and} $0 \leq Time < 20 \, min$, \textbf{then} $CONT$}
	
	\noindent \textit{\textbf{If} $0 \leq Dtemp < 0.25$ \textcelsius \ \textbf{and} $20 \leq Time < \, 25 \, min$, \textbf{then} $STOP$}
	
	\noindent \textit{\textbf{If} $0 \leq Dtemp < 0.25$ \textcelsius \ \textbf{and} $25 \leq Time < \, 30 \, min$, \textbf{then} $STOP$}
	
	\noindent \textit{\textbf{If} $0 \leq Dtemp < 0.25$ \textcelsius \ \textbf{and} $Time \geq \, 30 \, min$, \textbf{then} $STOP$}
	
	\noindent \textit{\textbf{If} $0.25\leq Dtemp < 0.35$ \textcelsius \ \textbf{and} $0 \leq Time < 20 \: min$, \textbf{then} $CONT$}
	
	\noindent \textit{\textbf{If} $0.25 \leq Dtemp < 0.35$ \textcelsius \ \textbf{and} $20 \leq Time < \: 25 min$, \textbf{then} $CONT$}
	
	\noindent \textit{\textbf{If} $0.25 \leq Dtemp < 0.35$ \textcelsius \ \textbf{and} $25 \leq Time < \, 30 \, min$, \textbf{then} $STOP$}
	
	\noindent \textit{\textbf{If} $0.25 \leq Dtemp < 0.35$ \textcelsius \ \textbf{and} $Time \geq \, 30 \, min$, \textbf{then} $STOP$}
	
	\noindent \textit{\textbf{If} $0.35\leq Dtemp < 0.5$ \textcelsius \ \textbf{and} $0 \leq Time < 20 \: min$, \textbf{then} $CONT$}
	
	\noindent \textit{\textbf{If} $0.35 \leq Dtemp < 0.5$ \textcelsius \ \textbf{and} $20 \leq Time < \: 25 min$, \textbf{then} $CONT$}
	
	\noindent \textit{\textbf{If} $0.35 \leq Dtemp < 0.5$ \textcelsius \ \textbf{and} $25 \leq Time < \, 30 \, min$, \textbf{then} $CONT$}
	
	\noindent \textit{\textbf{If} $0.35 \leq Dtemp < 0.5$ \textcelsius \ \textbf{and} $Time \geq \, 30 \, min$, \textbf{then} $STOP$}

	\noindent \textit{\textbf{If} $Dtemp \geq 0.5$ \textcelsius \ \textbf{and} $0 \leq Time < \: 20 min$, \textbf{then} $CONT$}
	
	\noindent \textit{\textbf{If} $Dtemp \geq 0.5$ \textcelsius \ \textbf{and} $20 \leq Time < \, 25 \, min$, \textbf{then} $CONT$}
	
	\noindent \textit{\textbf{If} $Dtemp \geq 0.5$ \textcelsius \ \textbf{and} $25 \leq Time < \, 30 \, min$, \textbf{then} $CONT$}
	
	\noindent \textit{\textbf{If} $Dtemp \geq 0.5$ \textcelsius \ \textbf{and} $Time \geq \, 30 \, min$, \textbf{then} $CONT$}
	
}
%\end{itemize*}
\begin{equation}
	\label{eqn:6}
\end{equation}

Now, considering the 6 rules in Equation~\ref{eqn:6} with stop testing outputs, then these 6 rules can be combined to become only 3 rules. The remaining 10 rules lead to continuing the testing procedure. Accordingly, Equation~\ref{eqn:6} is reduced to Equation~\ref{eqn:7}.\\

{\scriptsize
	
	\noindent \textit{\textbf{If} $0 \leq Dtemp < 0.25$ \textcelsius \ \textbf{and} $Time \geq \, 30 \, min$, \textbf{then} $STOP$}
	
	\noindent \textit{\textbf{If} $0.25 \leq Dtemp < 0.35$ \textcelsius \ \textbf{and} $Time \geq \, 30 \, min$, \textbf{then} $STOP$}
	
	\noindent \textit{\textbf{If} $0.35 \leq Dtemp < 0.5$ \textcelsius \ \textbf{and} $Time \geq \, 30 \, min$, \textbf{then} $STOP$}
	
	\noindent \textit{\textbf{Otherwise}, \textbf{then} $CONT$}
	
}
\begin{equation}
	\label{eqn:7}
\end{equation}

\section{The Fuzzy Inference Model Solution}
\label{sec:7}

\begin{figure}
	%\centering
	\includegraphics[width=0.4\textwidth]{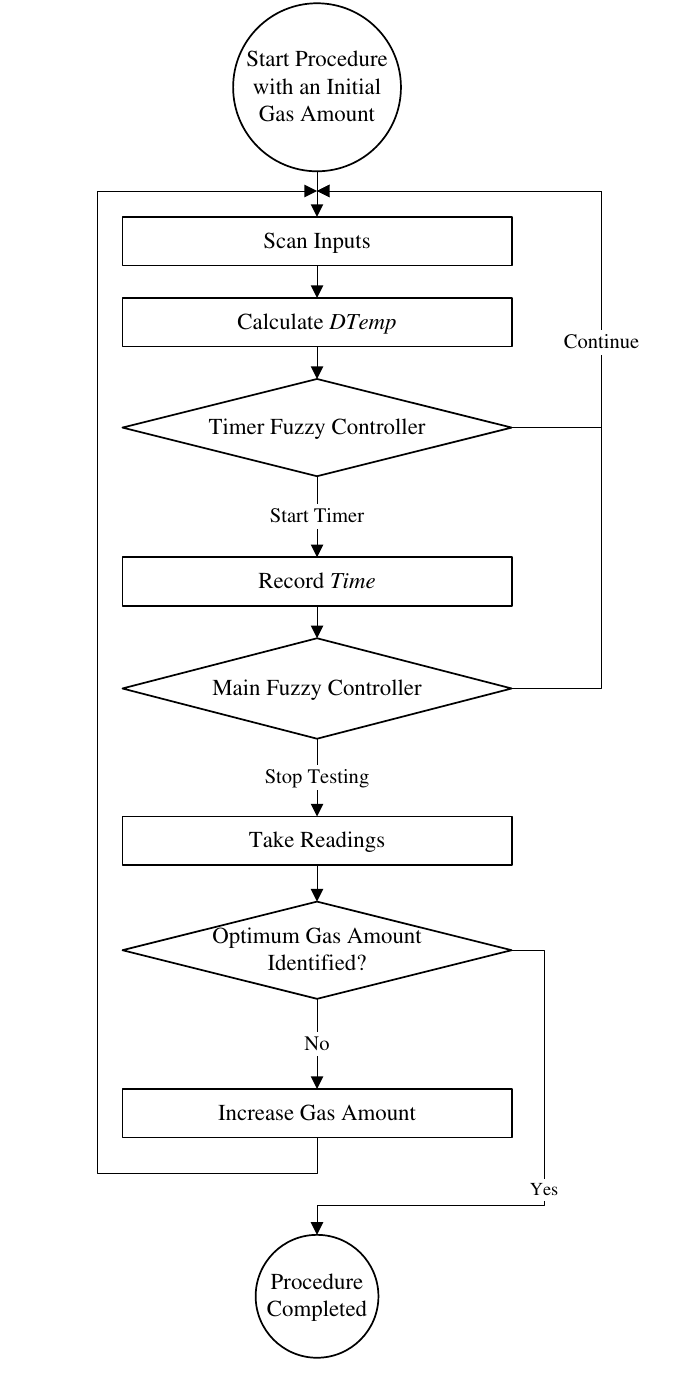}
	\caption{Refrigerant charging algorithm.}
	\label{fig:5}       % Give a unique label
\end{figure}

\begin{figure}
	%\centering
	\includegraphics[width=0.4\textwidth]{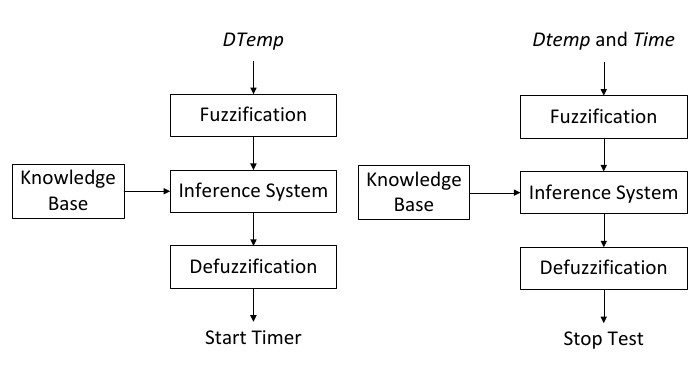}
	\caption{\textit{MISO} control configuration of the timer and main fuzzy controllers.}
	\label{fig:6}       % Give a unique label
\end{figure}

The fuzzy control model is formed by fuzzy sets and inference rules of the type given in Equation~\ref{eqn:1}, and thus reflecting the natural fuzziness of the problem. The model used in the testing procedure is depicted in Figure~\ref{fig:5} with all the necessary hardware for interfacing and measurements as explained in the sequel. The main fuzzy controller uses the Multiple Input Single Output (\textit{MISO}) control configuration shown in Figure~\ref{fig:6}. The controller has the two inputs $Dtemp$ and $Time$, and an output indicating the required charging action. Here, the (\textit{MFs}) of fuzzy sets are assigned over and representing a partitioning of these variables. The inference rules consist of all pair-wise combinations of these fuzzy sets, where each combination takes one set from each variable. The controller output is related to the stoppage or continuation of the testing procedure. 

%

The ranges of the controller variables, fuzzy sets, and the inference rules, are assigned based on consultations with experts in the refrigeration field. The consultations also helped improve the contents of Sections ~\ref{sec:3}, ~\ref{sec:5}, and ~\ref{sec:6}, which are also relied upon. Furthermore, some tuning has been done to the fuzzy sets and rules to adjust the model and make it, along with its results, more consistent with the expectations of the experts.

\begin{figure}
	%\centering
	\includegraphics[width=0.5\textwidth]{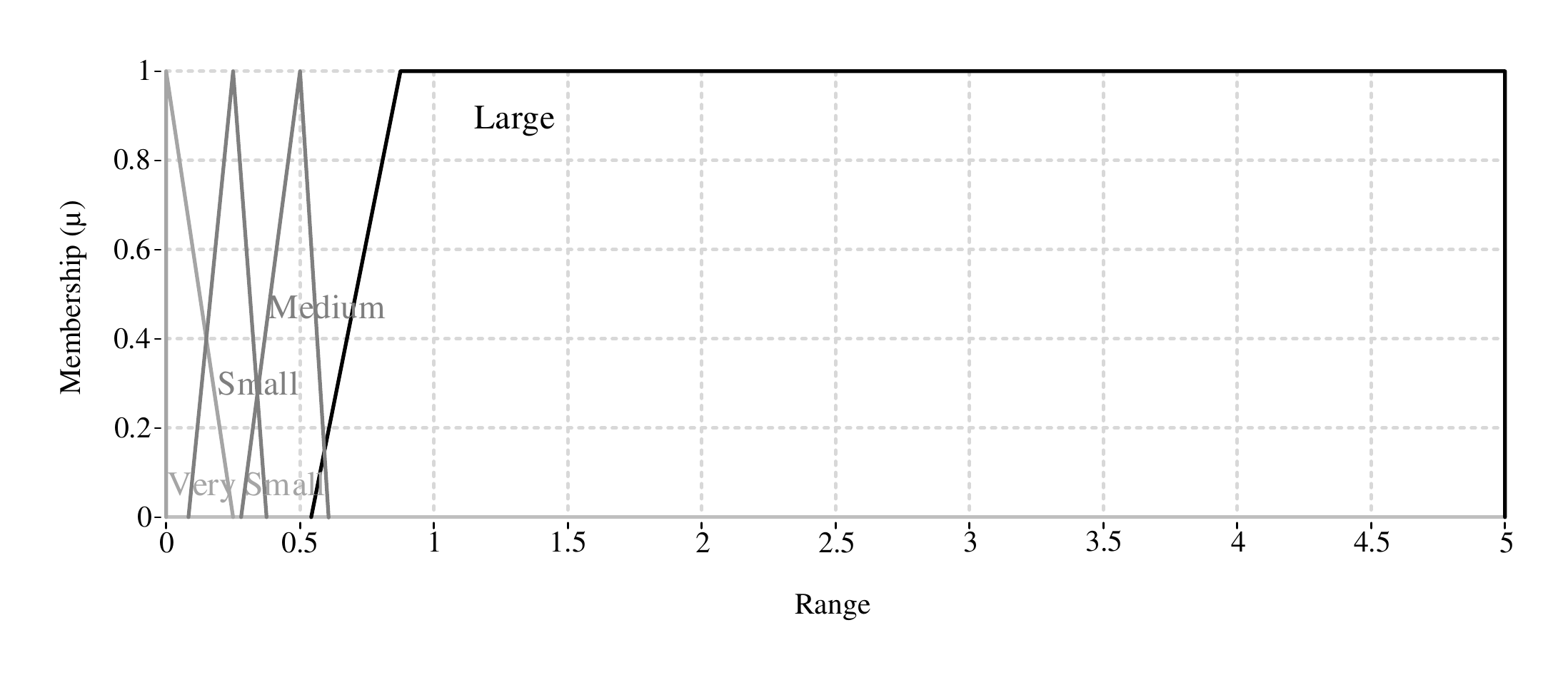}
	\caption{\added{\textit{MFs} of $Dtemp$.}}
	\label{fig:7}       % Give a unique label
\end{figure}

\begin{figure}
	%\centering
	\includegraphics[width=0.5\textwidth]{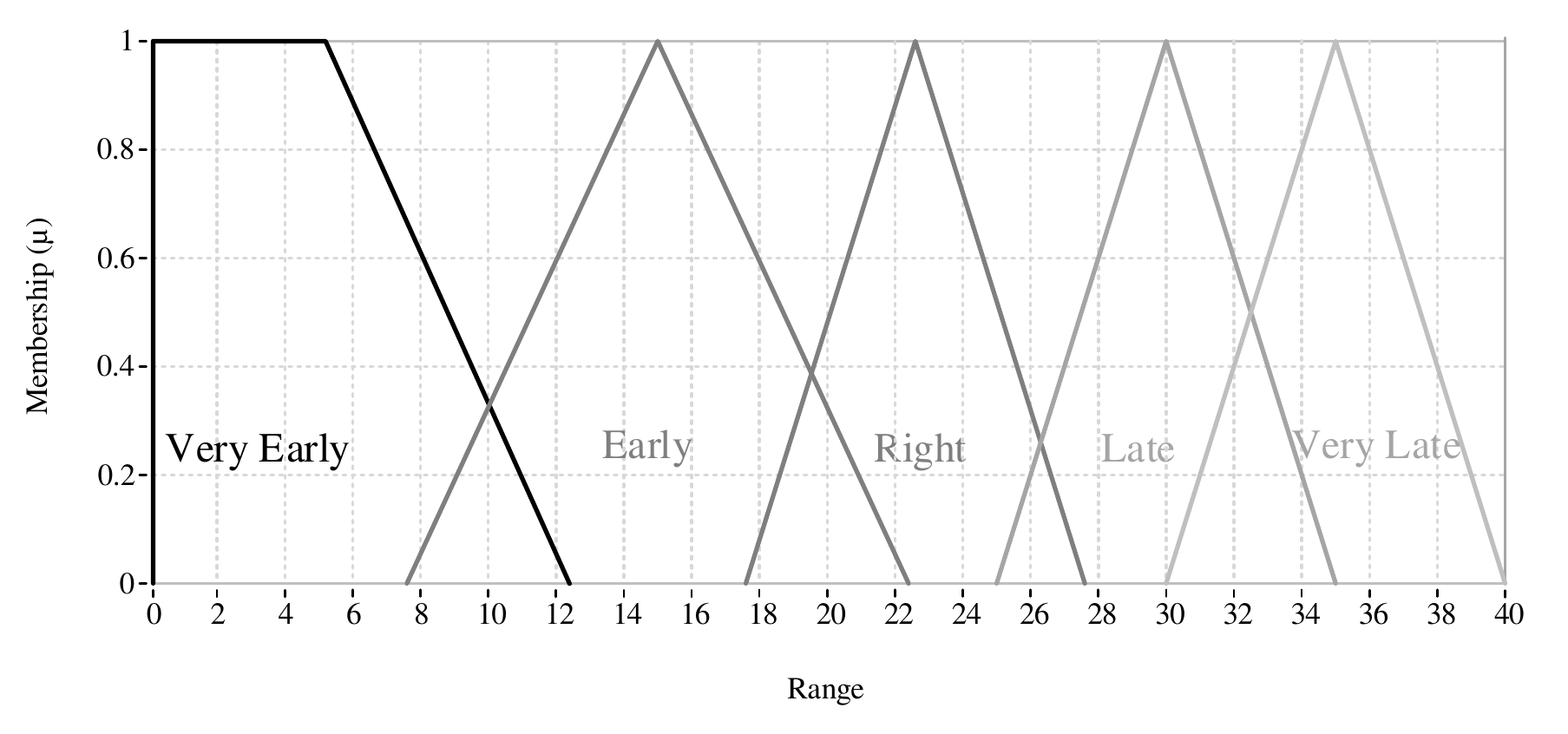}
	\caption{\added{\textit{MFs} of $Time$.}}
	\label{fig:8}       % Give a unique label
\end{figure}

\begin{figure}
	%\centering
	\includegraphics[width=0.5\textwidth]{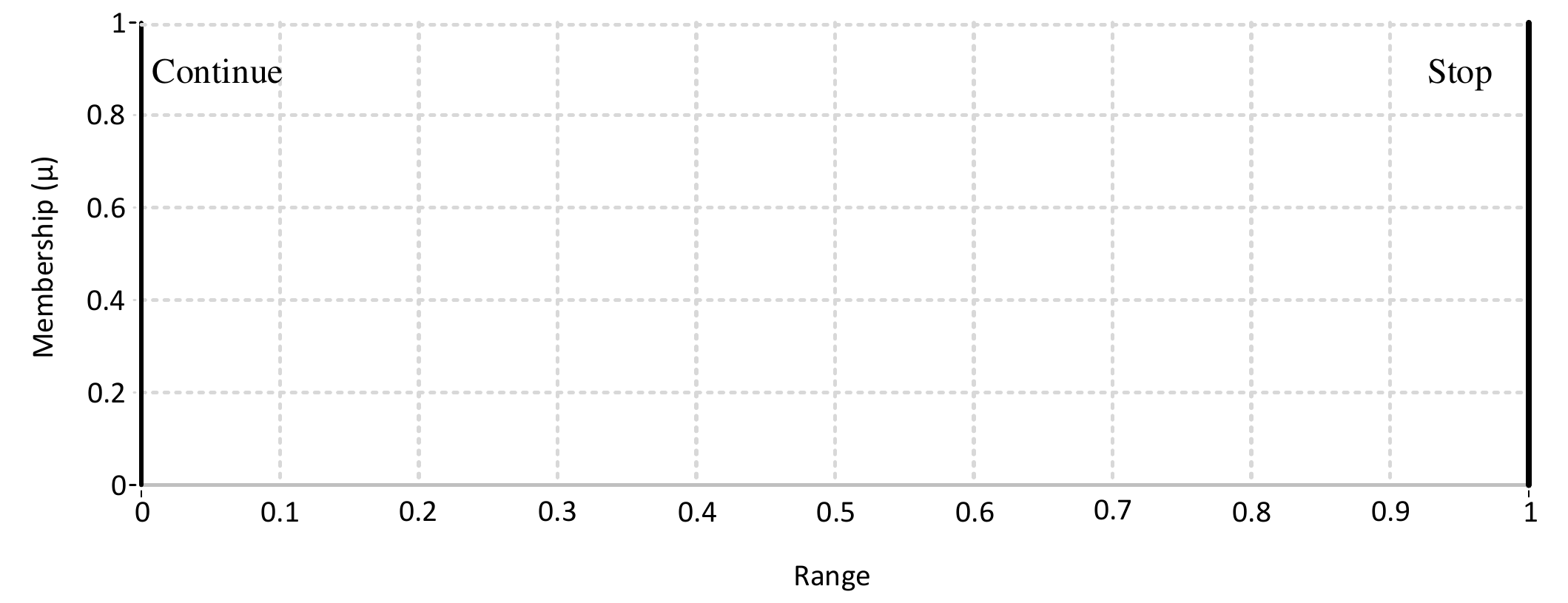}
	\caption{\added{\textit{MFs} of the main fuzzy controller’s output.}}
	\label{fig:9}       % Give a unique label
\end{figure}

The $Dtemp$ variable has a range between $0$ and $5$ {\textcelsius} and 4 fuzzy sets denoted as \textit{Very Small}, \textit{Small}, \textit{Medium} and \textit{Large} with \textit{MF}’s shown in Figure~\ref{fig:7}. As they relate to the problem description of Section~\ref{sec:3} and the linguistic terms used, the fuzzy sets used here can be interpreted as \textit{insignificant}, \textit{fairly insignificant}, \textit{significant} and \textit{highly significant} respectively. The $Time$ variable has a range between $0$ and $40$ minutes and the 5 fuzzy sets \textit{Very early}, \textit{Early}, \textit{Right}, \textit{Late}, and \textit{Very Late} with \textit{MF}’s shown in Figure~\ref{fig:8}. The linguistic interpretation of these fuzzy sets, as they relate to the time required to finish the testing process, can respectively be considered as \textit{insufficient}, \textit{fairly insufficient}, \textit{sufficient}, \textit{fairly sufficient} and \textit{highly sufficient}. The output variable of the fuzzy controller, denoted $Output$, has two singleton terms at $0$ and $1$ and they are respectively called $Stop$ and $Continue$ (See Figure~\ref{fig:9}). Based on the above partitioning, the main fuzzy controller uses the 20 inference rules given in Table~\ref{tbl:1}.

%

%

\begin{table*}
	
	\caption{Inference rules of the main controller.} 	
	\label{tbl:1}
	
	%\begin{center}
		\small
		
		\begin{tabular}{l | ccccc}	
			\hline 
			
			&  & & \textbf{\underline{\textit{Time}}} \\
			%\hline 
			
			\textbf{\textit{DTemp}} & \textbf{Very Early} & \textbf{Early} & \textbf{Right} & \textbf{Late} & \textbf{Very Late} \\			
			\hline
			\hline
			
			\textbf{Very Small} & Cont. & Stop & Stop & Stop & Stop \\
			\hline
			
			\textbf{Small} & Cont. & Cont. & Stop & Stop & Stop \\
			\hline
			
			\textbf{Medium} & Cont. & Cont. & Cont. & Stop & Stop \\
			\hline
			
			\textbf{Large} & Cont. & Cont. & Cont. & Cont. & Cont. \\
			\hline
			
		\end{tabular}
	%\end{center}
	
\end{table*}

\added{The choice of \textit{MFs} is of direct impact on the overall performance of the fuzzy system. Trapezoidal membership functions are formed using straightlines; thus, they are simple to capture and implement. Due to their smooth shape, Gaussian \textit{MFs} can provide appealing controller characteristics as they are non-zero at almost all points~\cite{ccaugman2013intuitionistic,6251210}.} A close examination of the inference rules’ structure reveals that the principle that should govern the intelligent behavior of a human expert is respected (See Section~\ref{sec:5}). In addition, the fuzzy sets \textit{MFs} \added{are} \deleted{have been} selected so that\added{,} when they are used with the rules in Table~\ref{tbl:1}, they would give an approximate stoppage time consistent with the expectations of the experts. \added{Here, the opinion of the expert plays the main role in the representation of \textit{MFs}.} If the rules in row 1 in Table~\ref{tbl:1} are considered, then these rules imply that for $Dtemp$ near $0.1$ {\textcelsius} (\textit{Very Small}), the stoppage time needs to be near 15 minutes (\textit{Early}). The rules in the second row imply that for $Dtemp$ near $0.25$ {\textcelsius} (\textit{Small}), then the stoppage time needs to be near 23 minutes (\textit{Right}). The third row imply that if $Dtemp$ is close to $0.5$ {\textcelsius} (\textit{Medium}), then the stoppage time needs to be close to 30 minutes (\textit{Late}). 

In fact, the 15, 23, and 30 minutes mentioned above are reasonable if the timing starts after having $Dtemp$ small enough as determined by the timer fuzzy controller whose rules can be set conveniently by the designer (See Equation~\ref{eqn:8}). Yet, the stoppage time instants can be changed easily and conveniently by tuning the fuzzy sets and rules to accommodate \added{for} various charging situations.

\begin{figure}
	%\centering
	\includegraphics[width=0.5\textwidth]{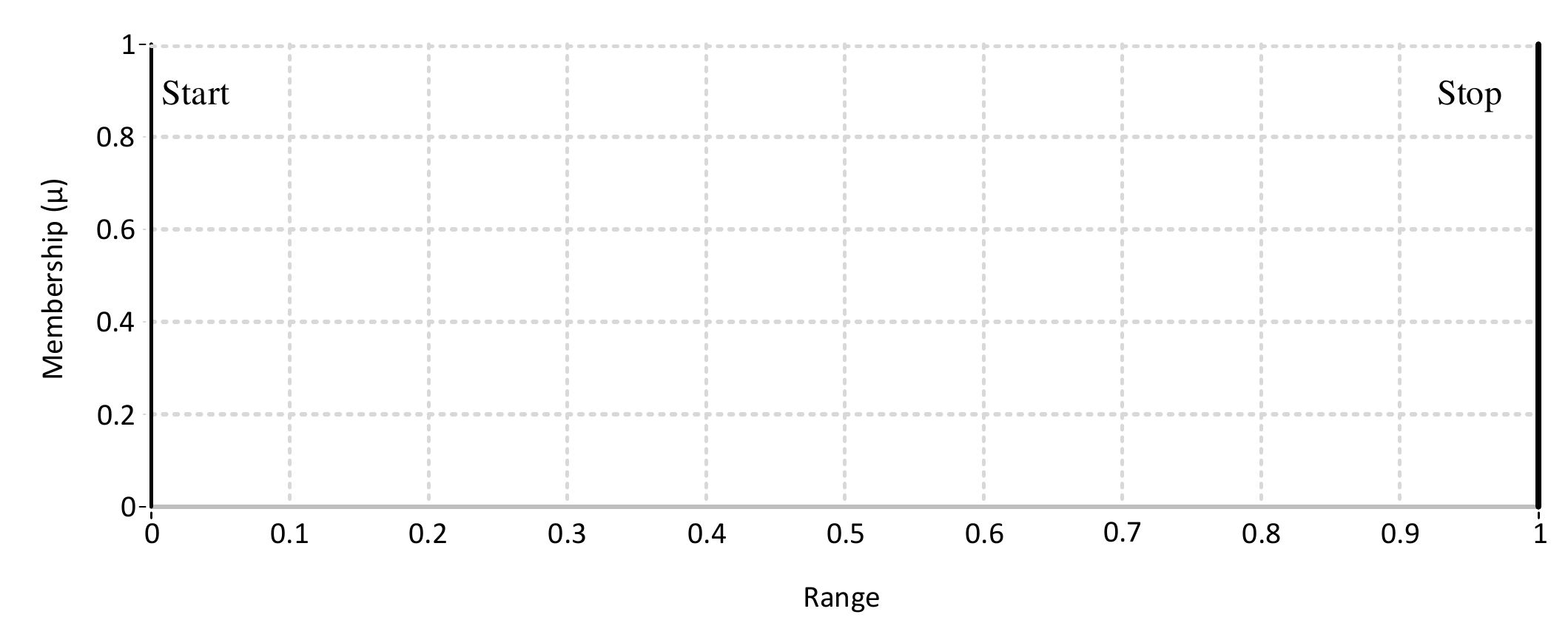}
	\caption{\added{\textit{MFs} of the timer fuzzy controller’s output.}}
	\label{fig:10}       % Give a unique label
\end{figure}

The Timer fuzzy controller has two variables; $Dtemp$ and $Output$. The two variables have the same \textit{MFs} as those in the main fuzzy controller. However, the output singletons are Stop and Start (See Figure~\ref{fig:10}). The inference rules of the timer controller are shown in Equation~\ref{eqn:8}.\\

{\scriptsize
	
	\noindent \textit{\textbf{If} $Dtemp \, is \, Very \, Small,$ \textbf{then} $Output \, is \, Start$}
	
	\noindent \textit{\textbf{If} $Dtemp \, is \, Small,$ \textbf{then} $Output \, is \, Start$}
	
	\noindent \textit{\textbf{If} $Dtemp \, is \, Medium,$ \textbf{then} $Output \, is \, Start$}
	
	\noindent \textit{\textbf{If} $Dtemp \, is \, Large,$ \textbf{then} $Output \, is \, Stop$}
	
}
\begin{equation}
	\label{eqn:8}
\end{equation}

\begin{figure}
	%\centering
	\includegraphics[width=0.5\textwidth]{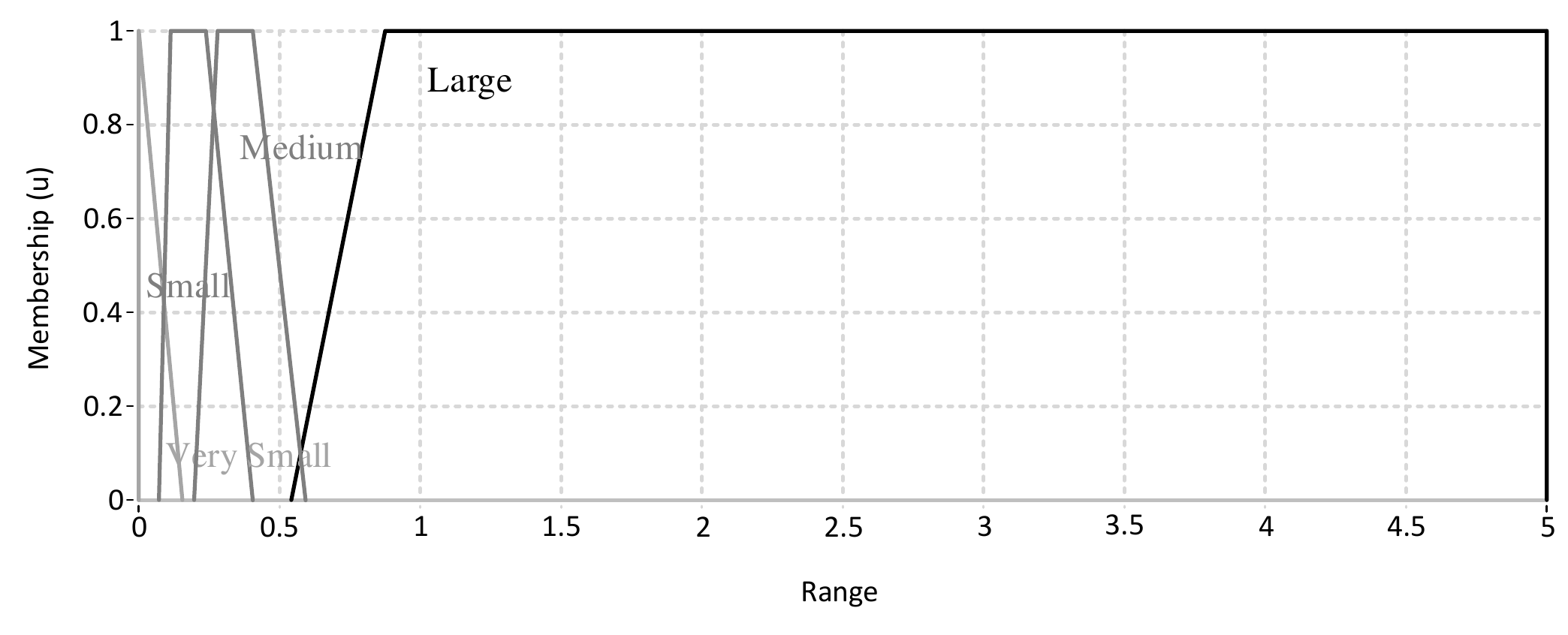}
	\caption{\added{Trapezoidal \textit{MFs} of \textit{Dtemp}.}}
	\label{fig:11}       % Give a unique label
\end{figure}

\begin{figure}
	%\centering
	\includegraphics[width=0.5\textwidth]{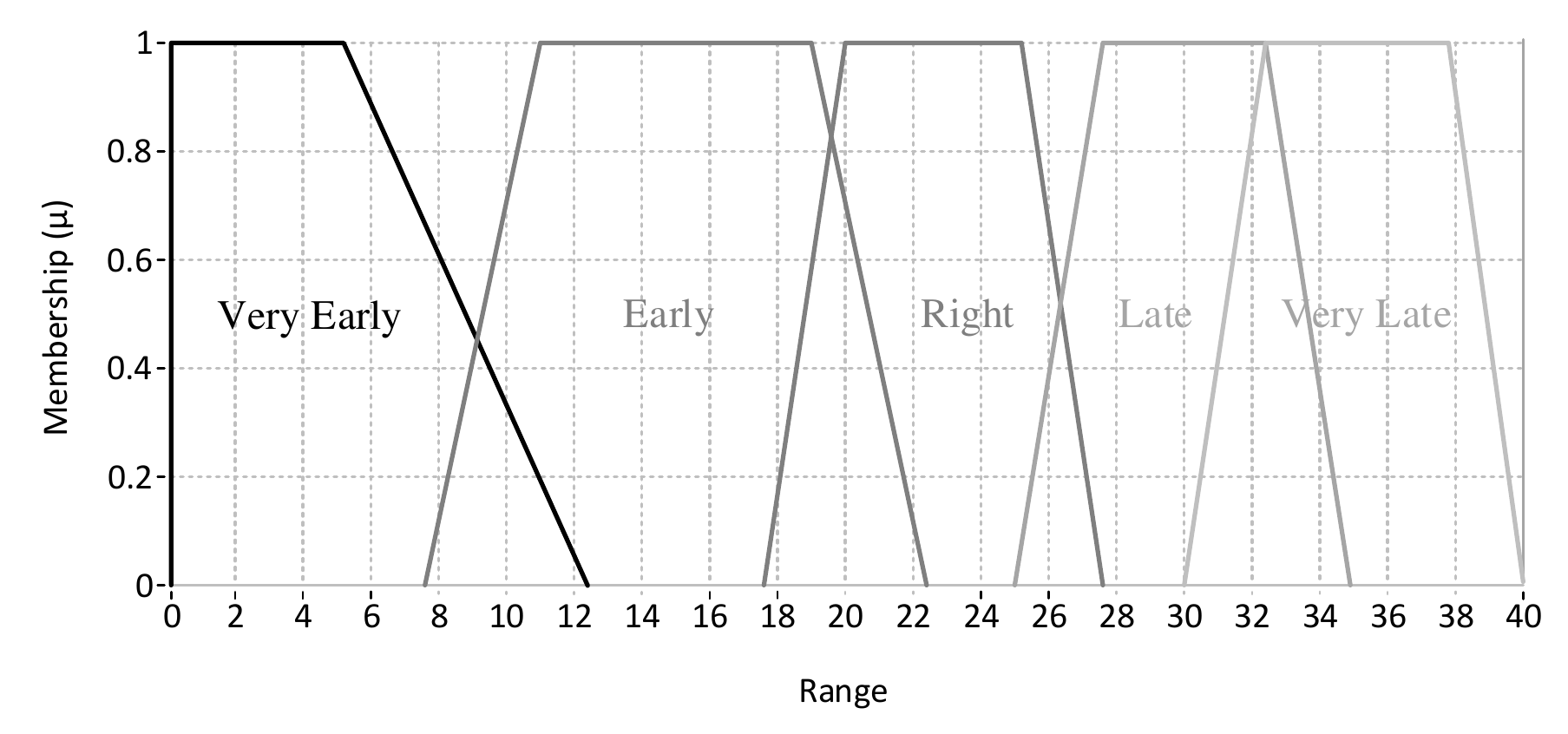}
	\caption{\added{Trapezoidal \textit{MFs} of \textit{Time}.}}
	\label{fig:12}       % Give a unique label
\end{figure}

\begin{figure}
	%\centering
	\includegraphics[width=0.5\textwidth]{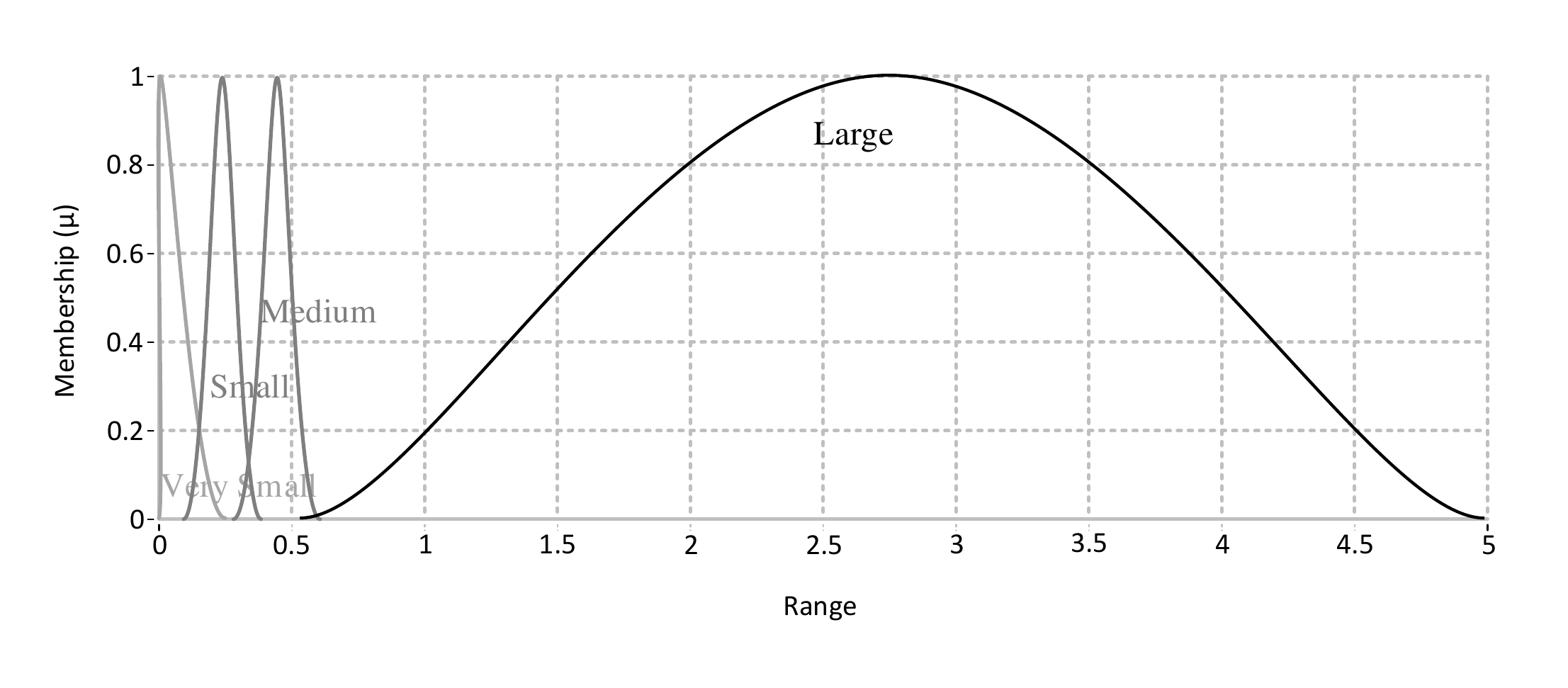}
	\caption{\added{Gaussian \textit{MFs} of \textit{Dtemp}.}}
	\label{fig:13}       % Give a unique label
\end{figure}

\begin{figure}
	%\centering
	\includegraphics[width=0.5\textwidth]{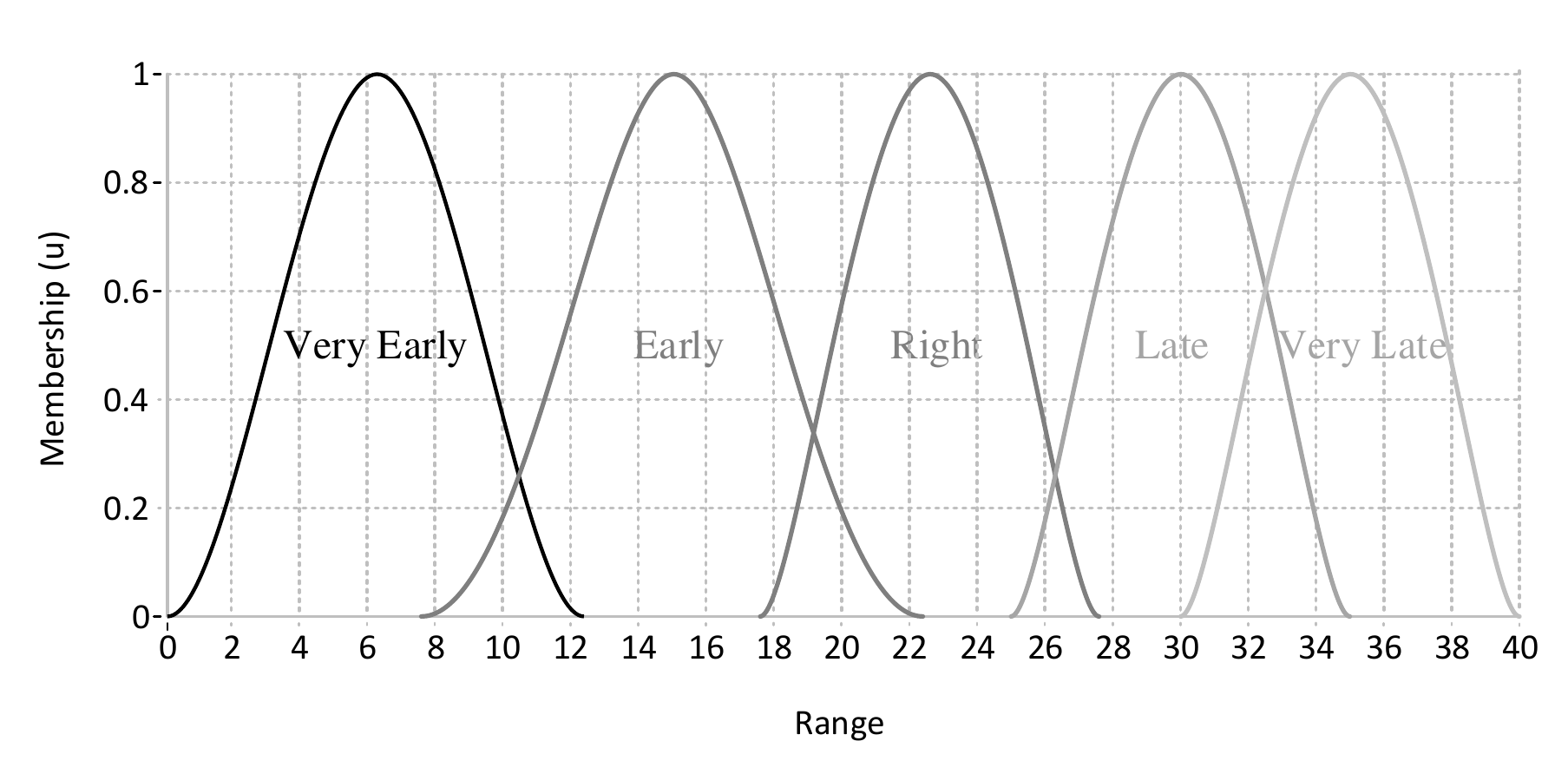}
	\caption{\added{Gaussian \textit{MFs} of \textit{Time}.}}
	\label{fig:14}       % Give a unique label
\end{figure}

A combination triangular and trapezoidal MFs are used for the representation of $DTemp$ and $Time$ as in Figures~\ref{fig:7} and~\ref{fig:8}. However, fully-trapezoidal and Gaussian-only alternatives are included in the investigation (See Figures~\ref{fig:11},~\ref{fig:12},~\ref{fig:13}, and~\ref{fig:14}).

%

%

%

\section{System Architecture and Implementation}
\label{sec:8}

The algorithm in Figure~\ref{fig:5} is implemented as part of a computerized \textit{DAQ}, monitoring and decision-making system used for refrigerant charging, testing and performance evaluation of refrigerators. The system is implemented under LabVIEW using the available Fuzzy Logic Toolkit and contains a complete hardware system for interfacing and measurement. Using the Fuzzy Logic Toolkit functions, the design and tuning of the fuzzy logic controller, as well as the activation of the inference engine for real-time process control, can be easily done. 

Two software processes, a scanner and an analyzer, are included in the testing procedure shown in Figure~\ref{fig:5}. The scanning process acquires the temperature readings from the thermocouples connected to the hardware interfacing cards and controls the reading speed. The analyzer processes the data acquired by the scanner. In addition, the analyzer monitors the changes in temperature reading and records the lowest attained temperature once the main controller decides to stop the testing for a specific refrigerant amount. As the charging is repeated, until the optimum amount is identified, the analyzer draws a chart of refrigerant amount versus temperature as in Figure~\ref{fig:4}. 

\begin{figure}
	%\centering
	\includegraphics[width=0.5\textwidth]{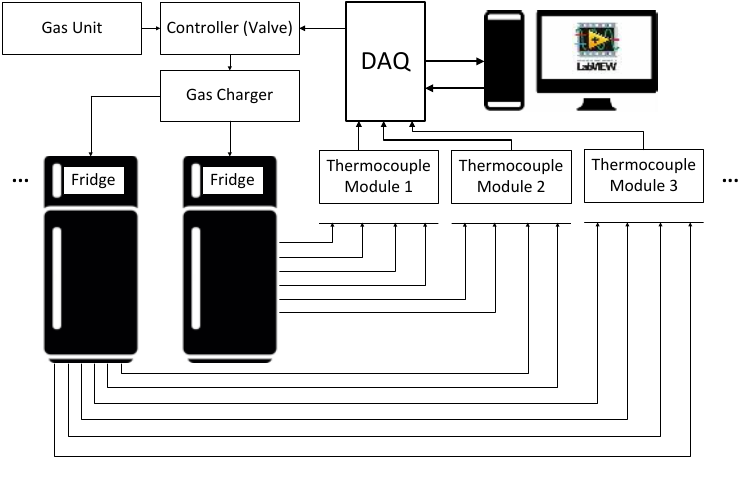}
	\caption{Gas Charging System Architecture. Two fridges out of eight are shown.}
	\label{fig:15}       % Give a unique label
\end{figure}

The model illustrated in Figure~\ref{fig:15} consists of two out of eight fridges being charged in parallel. The NIcDAQ-9174 \textit{DAQ} Chassis is used along with three NI 9211 Thermocouple Modules. Six thermocouples are used for each fridge at various locations to guarantee the accuracy of the process. After acquiring the required data from the thermocouples attached to the three NI 9211 Thermocouple Modules, it gets processed by the developed software to determine $Dtemp$ and accordingly the $Time$, $Start$, $Continue$, and $Stop$ of the timer and the refrigerant charging along with the amount of gas required. \added{The refrigerant charging unit is controlled using normally-closed solenoid valves and NI DAQ C-Series Relay Output Modules.}

\begin{figure*}
	%\centering
	\includegraphics[width=0.95\textwidth]{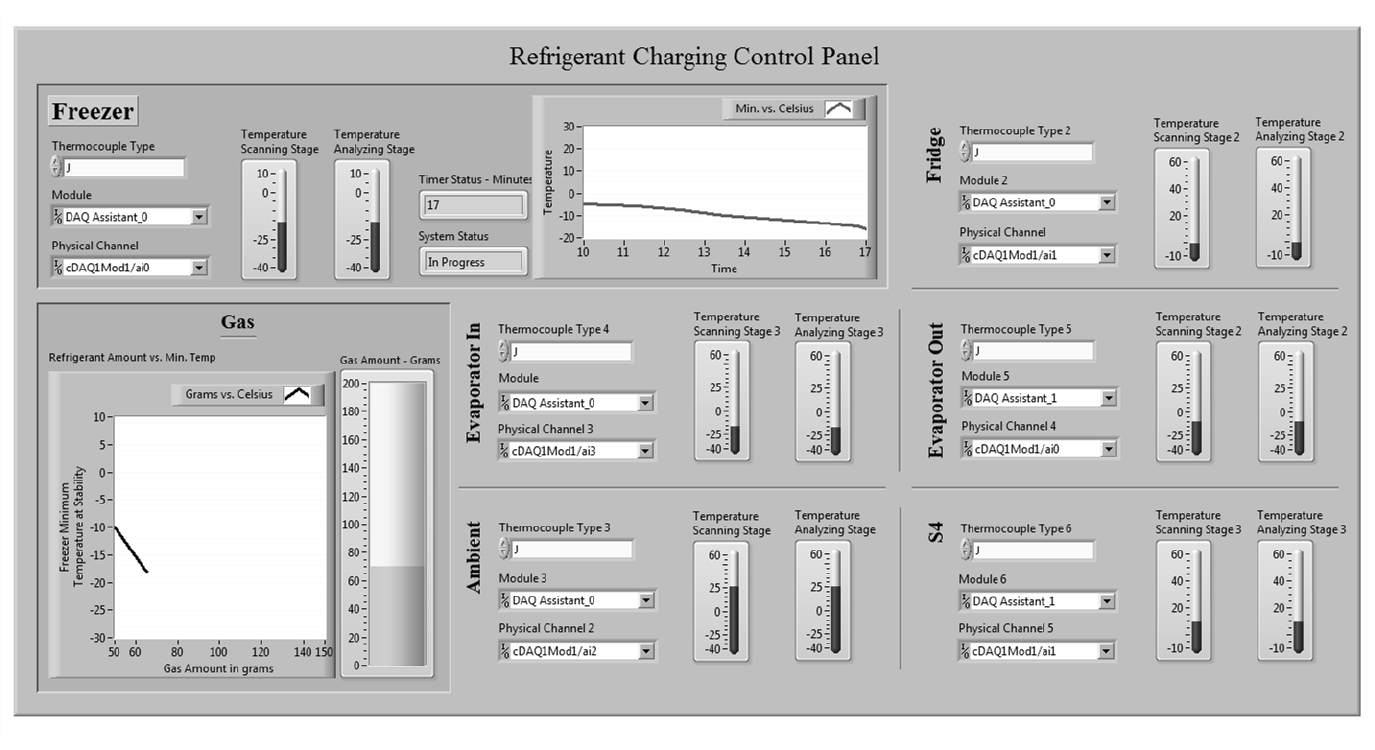}
	\caption{The refrigerant charging main control panel.}
	\label{fig:16}       % Give a unique label
\end{figure*}

The main elements of the \textit{GUI} include the freezer and gas controllers (See Figure~\ref{fig:16}). The \textit{GUI} also includes monitors for the temperature change at different parts of the refrigeration system. The temperature monitoring for every part of the system has thermocouple type, module, and, physical channel list selectors. The module list selector is used for choosing the designated \textit{DAQ} module as several modules are mounted on each \textit{DAQ}. The physical channel specifies one out of eight possible thermocouples, per \textit{DAQ}, to monitor. Several temperature indicators are added for each stage of monitoring (scanning and analyzing). 

The scanning and analyzing using the fuzzy controllers are applied on the freezer part of the refrigerator. The freezer monitoring window includes a state indicator that reflects the status of the procedure while analyzing and can be either \textit{in progress} or \textit{stopped}. The freezer window also includes a temperature waveform charts that plots the temperature versus time. Lastly, a timer in minutes is added to the freezer’s window for noting the time the process is taking.

\begin{figure*}
%	\centering
	\includegraphics[width=0.95\textwidth]{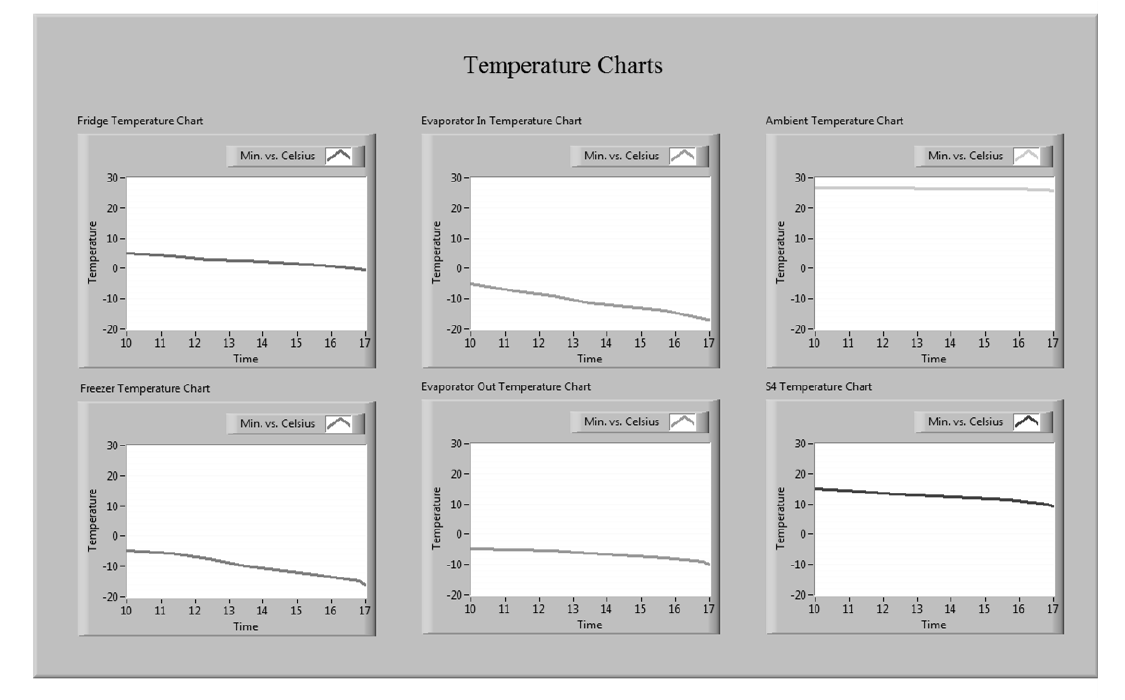}
	\caption{The refrigerant charging temperature monitoring charts.}
	\label{fig:17}       % Give a unique label
\end{figure*}

The gas monitoring window comprises a quantity indicator, and a waveform chart that depicts the freezer’s minimum temperature at stability versus the gas amount in grams. Moreover, the user inter-face includes waveform charts to monitor the state of the temperature throughout the stages and visually asserting whether stability has been reached or not (See Figure~\ref{fig:17}). For one unit, six thermo-couples are distributed among the ambient, fridge, freezer, evaporator-in, evaporator-out, and the suction tube (\textit{S4}). As the application software supports eight units, the user has the option to hide parts of the \textit{GUI} window if a smaller number of units are being tested.

\section{Analysis and Evaluation}
\label{sec:9}

\subsection{Results}

Based on the \textit{MFs} in Figure~\ref{fig:7}, the rules in Equation~\ref{eqn:8}, and Equation~\ref{eqn:5}, the output of the timer fuzzy controller changes per $Dtemp$ values. As the values change between the specified margins, the degree of belongingness to each of the Continue or Stop fuzzy sets changes accordingly. Setting a threshold value somewhere between $0.5$ {\textcelsius} and $0.75$ {\textcelsius} to start the timer shows the acquired ability to modify the start time in accordance with the way by which the human expert would perceive the $Dtemp$ values as indicators for reaching the stability region. If necessary, the range of $Dtemp$ threshold can be modified as well by varying the rules in Equation~\ref{sec:8}. Thus, the value of $Dtemp$ at which the timer starts can be adjusted in accordance with the human expertise.

\begin{figure}
%	\centering
	\includegraphics[width=0.5\textwidth]{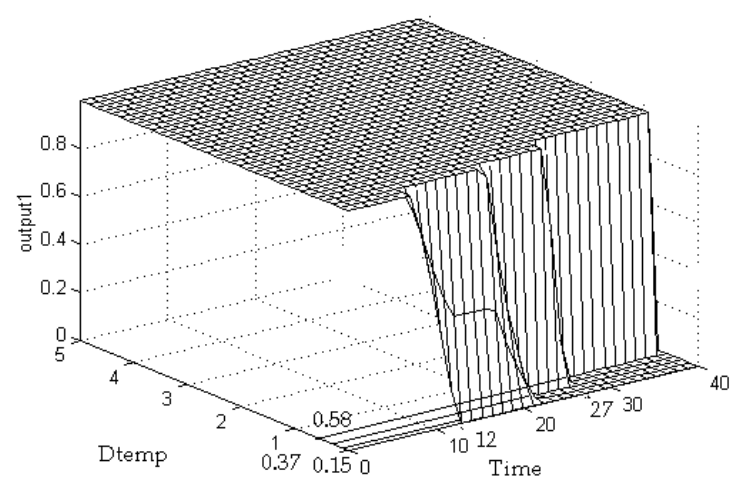}
	\caption{The output of the main fuzzy controller.}
	\label{fig:18}       % Give a unique label
\end{figure}

\deleted{Based on} \added{According to} the \textit{MFs} in Figure~\ref{fig:7} and Figure~\ref{fig:8}, the rules in Table~\ref{tbl:1}, and Equation~\ref{eqn:7}, Figure~\ref{fig:18} shows the main controller’s output versus $Dtemp$. Thus, if $Dtemp$ is in the range $[0$ {\textcelsius}$, \, 0.15$ {\textcelsius}$]$, then the controller stops the test in 12 minutes. Now, if $Dtemp$ is between $(0.15$ {\textcelsius}$, \, 0.37$ {\textcelsius}$]$, then the controller stops the testing in 20 minutes. For $Dtemp$ in $(0.37$  {\textcelsius}$, \, 0.58$ {\textcelsius}$]$, the controller waits more time to stop the test (27 minutes). Moreover, if the value of $Dtemp$ is between $(0.58$ {\textcelsius}$, \, 5$ {\textcelsius}$]$, then the controller keeps the test running. These results are reasonable in terms of their compatibility with the expert’s expectations as considered while setting the membership functions and rules in Section~\ref{sec:7}.

The plot of the main-controller output versus $Dtemp$ and $Time$ (Figure~\ref{fig:18}) shows that when $Dtemp$ becomes smaller, then the observation period is also made smaller permitting an earlier stoppage time instant. In addition, the 12, 20 and 27 minutes are not the only time values at which the testing can stop. When the timer starts, $Dtemp$ values could, for example, be in the range $[0.37$ {\textcelsius}$, \, 0.58$ {\textcelsius}$]$ leading to an anticipated stoppage time at 27 minutes. But, as the time progresses these values of $Dtemp$ could decrease and become in $[0.15$ {\textcelsius}$, \, 0.37$ {\textcelsius}$]$ and then in $[0$ {\textcelsius}$, \, 0.15$ {\textcelsius}$]$. When \deleted{this} \added{such a decrease} happens, the time could have reached any value between 12 and 20 minutes at which the testing stops. Also, the passage of $Dtemp$ values from the first range to the second could occur at any time between 20 and 27 minutes.  \deleted{So} \added{To that end}, the stoppage time of the test can be at any value between 12 and 27 minutes. Hence, the continuity aspect, which can be attributed to human behavior especially in the control area, is obtained here.

\added{To evaluate the performance of the proposed fuzzy control algorithms we develop the following indicators~\cite{DAMAJ2017,7870083,patterson2017computer}:}
	{\small
	\begin{enumerate}
		\item \added{\textbf{Test Time (\textit{TT})}: The time, in Minutes (min), taken to complete the test of one unit.}
		\item \added{\textbf{Time Saving (\textit{TS})}: The difference between \textit{TTs} of two controllers.}
		\item \added{\textbf{Throughput (\textit{TH})}: The number of tests per unit time in Tests per Hour (\textit{TpH}) or Tests per a 12-hour work Day (\textit{TpD}).}
		\item \added{\textbf{Performance (\textit{PR})}: The \textit{PR} of a controller is inversely proportional to its \textit{TT}.}
		
		\newcounter{enumTemp}
		\setcounter{enumTemp}{\theenumi}
\end{enumerate}}

\noindent {\small \added{The \textit{PR} indicator is used to calculate Performance Ratios ($PR_{ratio}$)~\cite{patterson2017computer} according to Equations~\ref{eqn:PR} and ~\ref{eqn:PRratio}. The \textit{PR} indicator enables the calculation of the percentage performance improvement of one controller over the other.}}

\begin{equation}
\label{eqn:PR}
PR_{ratio} = \dfrac{PR_{controller_{1}}}{PR_{controller_{2}}}
\end{equation}

\noindent where $PR_{ratio}$ is in $\%$, and

\begin{equation}
\label{eqn:PRratio}
\dfrac{PR_{controller_{1}}}{PR_{controller_{2}}} = \dfrac{TT_{controller_{2}}}{TT_{controller_{1}}}
\end{equation}

{\small
	\begin{enumerate}
		\setcounter{enumi}{\theenumTemp}
		
		\item \added{\textbf{Energy Consumption (\textit{E})}: in kilo Watts per hour (\textit{kWh}).}
\end{enumerate}}

With an aim to experiment test time savings, different membership functions are investigated. Table~\ref{tbl:2}, presents the implementation results of triangular, trapezoidal and Gaussian \textit{MFs}. From Table~\ref{tbl:2}, it can be clearly seen that the Fuzzy controllers outperformed the crisp controllers. Nevertheless, minimal performance differences are detected between different types of membership functions for the developed fuzzy controller. The trapezoidal function provides the best performance for the designated $Dtemp$ ranges with a \added{70.6\% performance improvement based-on the \textit{$PR_{ratio}$} and a maximum \textit{TS} of 21.2 min,} followed by the Gaussian with 60\%, then the triangular with 36.6\%. The performance difference is due to the different shapes of the \textit{MFs} which affects the calculation of the degree of membership to a certain fuzzy set. \added{In the presented \textit{TT} measurements, the valve dynamics can be ignored relevant to the dominant dynamics of the slow temperature process. With a maximum close or open time of 1 Second, and the small amount of gas charge increment, the solenoid valve operation and the refrigerant charging introduce no significant time delays to the overall process~\cite{Honeywell18}.}    

\begin{table*} 		
%	\begin{center}
		\caption{\added{The measured \textit{TT} at different values of $Dtemp$ with the \textit{TS} and \textit{$PR_{ratio}$} of the different controller implementations over the crisp controller of Equation~\ref{eqn:5}.}} 
		\label{tbl:2}
		%\tiny
		
		\begin{tabular}{c || c|c|c|c|c}	
			\hline 
			
			&  & & \textbf{\underline{Controllers}} \\
			
			& \textbf{Crisp} & \textbf{Crisp}& \textbf{Fuzzy} & \textbf{Fuzzy}& \textbf{Fuzzy} \\			
			
			& \textbf{Equation~\ref{eqn:5}} & \textbf{Equation~\ref{eqn:7}}& \textbf{Triangular} & \textbf{Trapezoidal}& \textbf{Guassian} \\			
			
			\textbf{\textit{DTemp Range}} & & & \textbf{MFs} & \textbf{MFs}& \textbf{MFs} \\			
			\hline
			\hline
			
			%%%%%%%%%%%%%%%%%%%%%%%%%%%
			$0.4 \leq DTemp \leq 0.5$ & \textbf{30 min} & \textbf{30 min} & \textbf{27 min} & \textbf{[26, 27] min} & \textbf{26.2 min} \\
			\hline
			
			{\scriptsize \textit{TS}} & {\scriptsize-} & {\scriptsize0 min} & {\scriptsize3 min} & {\scriptsize[4, 3] min} & {\scriptsize3.8 min}\\
			\hline		

			{\scriptsize\textit{$\%$ Improvement based-on $PR_{ratio}$}} & 	{\scriptsize -} & 	{\scriptsize 0\%} & 	{\scriptsize 10\%} & 	{\scriptsize [13\%, 10\%]} & 	{\scriptsize 12.6\%}\\
			\hline		
			\hline
			
			%%%%%%%%%%%%%%%%%%%%%%%%%%%%%%%%%%
			$0.35 \leq DTemp \leq 0.4$ & \textbf{30 min} & \textbf{30 min} & \textbf{[25, 27] min} & \textbf{[26, 27] min} & \textbf{26.2 min} \\
			\hline
			
			{\scriptsize \textit{TS}} & {\scriptsize -} & {\scriptsize 0 min} & {\scriptsize [5, 3] min} & {\scriptsize [4, 3] min} & {\scriptsize 3.8 min}	\\		
			\hline
			
			{\scriptsize \textit{$\%$ Improvement based-on $PR_{ratio}$}} & {\scriptsize -} & {\scriptsize 0\%} & {\scriptsize [13\%, 10\%]} & {\scriptsize [13\%, 10\%]} & {\scriptsize 12.6\%}	\\		
			\hline
			\hline
			
			%%%%%%%%%%%%%%%%%%%%%%%%%%%%%%%%%%
			
			$0.25 \leq DTemp \leq 0.3$ & \textbf{30 min} & \textbf{25 min} & \textbf{20 min} & \textbf{[19.1, 20] min} & \textbf{19.3 min} \\
			\hline
			
			{\scriptsize \textit{TS}} & {\scriptsize -} & {\scriptsize 5 min} & {\scriptsize 10 min} & {\scriptsize [10.9, 10] min} & {\scriptsize 10.7 min}\\
			\hline			
			
			{\scriptsize \textit{$\%$ Improvement based-on $PR_{ratio}$}} & {\scriptsize -} & {\scriptsize 16.7\%} & {\scriptsize 33.3\%} & {\scriptsize [33.6\%, 33.3\%]} & {\scriptsize 35.6\%}\\
			\hline
			\hline			
			
			%%%%%%%%%%%%%%%%%%%%%%%%%%%%%%%%%%
			
			$0 \leq DTemp \leq 0.25$ & \textbf{30 min} & \textbf{20 min} & \textbf{[12, 20] min} & \textbf{[8.8, 19.1] min} & \textbf{[19.1, 19.3] min}\\
			\hline

			{\scriptsize \textit{TS}} & {\scriptsize -} & {\scriptsize 10 min} & {\scriptsize [18, 10] min} & {\scriptsize [21.2, 10.9] min} & {\scriptsize [10.9, 10.7] min}\\	
			
			{\scriptsize \textit{$\%$ Improvement based-on $PR_{ratio}$}} & {\scriptsize -} & {\scriptsize 33\%} & {\scriptsize [60\%, 33.3\%]} & {\scriptsize [70.6\%, 36.3\%]} & {\scriptsize [36.3\%, 35.6\%]}\\			
			\hline
			
		\end{tabular}
%	\end{center}	
\end{table*}

With no doubt, the time saving leads to significant reduction in energy consumption of the testing process. The energy saving is proportional to the time saving given Equation~\ref{eqn:9} and can reach a maximum of 70.6\% saving per unit using the fuzzy controller with trapezoidal \textit{MFs}. We consider that a modern refrigerator unit consumes a total power (\textit{P}) of 2.5 \textit{kW} \cite{samadi2013tackling}.

\begin{equation}
	\label{eqn:9}
	E \, = P \, \times t 
\end{equation}

\noindent Where, \textit{t} is the time in hours, \textit{E} in \textit{kWh}, and \textit{P} in \textit{kW}\\

\begin{figure}
%	\centering
	\includegraphics[width=0.5\textwidth]{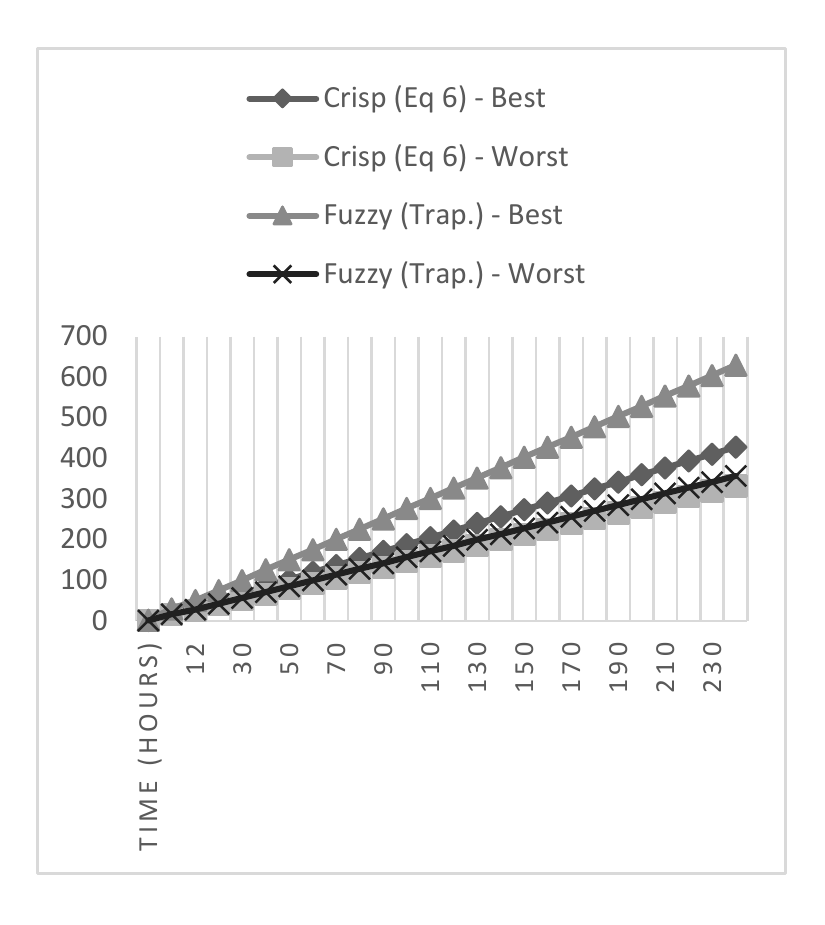}
	\caption{The number of tests versus time for one unit.}
	\label{fig:19}       % Give a unique label
\end{figure}

Figure~\ref{fig:19} shows the number of tests per time that can be run versus time$\textendash$given a maximum setup time of 15 minutes per test. Here, a test means that the minimum temperature has been detected once a new amount of refrigerant has been added. Usually, to decide on the best gas amount, 4-6 tests are required. The figure considers the worst and best calculated time savings for the crisp and fuzzy controllers. The Figure shows that the crisp controller of Equation~\ref{eqn:6} performed almost equally to the fuzzy controller with trapezoidal \textit{MFs} but at its worst-case duration of test. The fuzzy controller with trapezoidal \textit{MFs} outperformed all other models in its best-case time saving of 70.6\%.

\begin{figure}
%	\centering
	\includegraphics[width=0.5\textwidth]{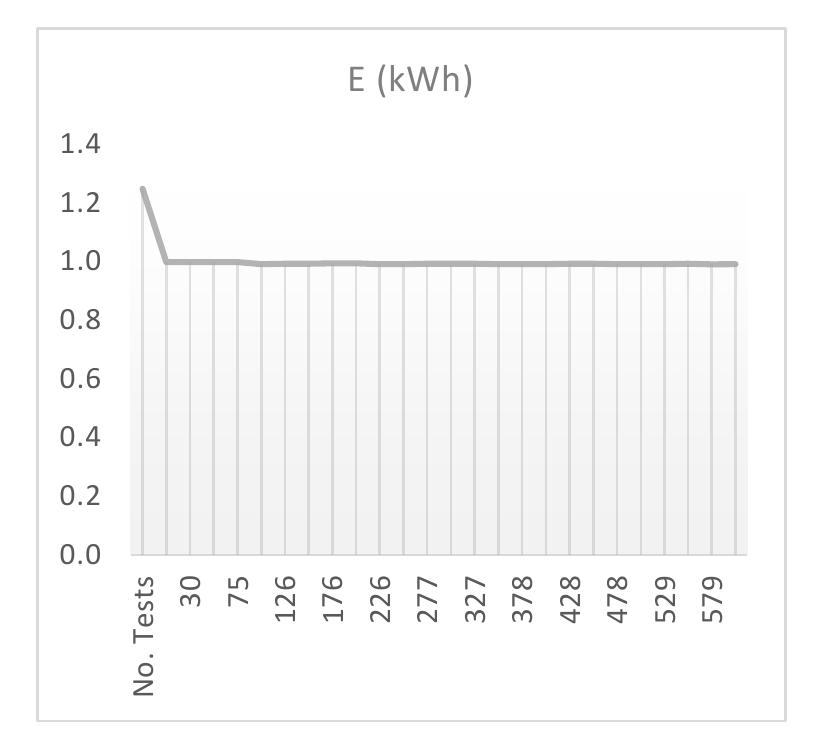}
	\caption{The energy consumption per test versus time for one unit and using the fuzzy controller with trapezoidal \textit{MFs} at the best time saving of 70.6\%.}
	\label{fig:20}       % Give a unique label
\end{figure}

Table~\ref{tbl:3} and Figure~\ref{fig:20} show the calculated energy consumption per test versus time. The Table considers the worst and best test times for both the crisp and fuzzy controllers. \added{The results in the Table confirms that the Trapezoidal \textit{MFs} enable the highest \textit{TH} of 30 \textit{TpD}}. The Figure shows that the energy consumption per test per unit is almost constant over time with a value of 1 kWh. The results shown in the table are for the controller with trapezoidal \textit{MFs} at its best-case test time.

\begin{table*} 	
%	\begin{center}
		\caption{Energy consumption per test for the worst and best times for both of crisp and fuzzy controllers. The maximum test setup time is considered 15 minutes, while the power consumption is 2.5 kW. The results shown are for one unit.} 
		\small
		\label{tbl:3}
		
		\begin{tabular}{p{5.3 cm}| p{3.5 cm}  | p{3.5 cm} | p{3.2 cm}}	
			
			\hline 
			\textbf{Controller} & \textbf{Test Time} in minutes & \textbf{No. of Tests} in 12 Hours &\textbf{E per Test} kWh/test\\
			
			\hline \hline
			Crisp (Equation~\ref{eqn:7}) - \textbf{Best} & 20  &  20 & 1.5\\
			
			\hline 
			Crisp (Equation~\ref{eqn:7}) - \textbf{Worst} & 30 & 16 & 1.87\\
			
			\hline 
			Fuzzy (Trapezoidal \textit{MFs}) - \textbf{Best} & 8.8 & 30 & 1\\
			
			\hline
			
			Fuzzy (Trapezoidal \textit{MFs}) - \textbf{Worst} & 27 & 17 & 1.76\\
			
			\hline
			
		\end{tabular}		
%	\end{center}	
\end{table*}

\subsection{General Evaluation}
In addition to the above-mentioned results, provided by the fuzzy models, Table~\ref{tbl:4} contains comparisons among the fuzzy model of Table~\ref{tbl:1} and the classical inference model of Equation~\ref{eqn:7}. The main differences between the two models are the complexity of required procedural support to the software tool, and the complexity of the development and tuning of the model. The main complexity is caused by the procedures which need to be used to make the crisp-controller results consistent with the experts’ expectations (See Section~\ref{sec:5}). The consistency can be achieved by either adjusting the limits of the adopted intervals partitioning $Dtemp$ and $Time$ variables, or by increasing the number of these intervals. In the first case, however, and since the results depend on the limits of the intervals, then comparatively more effort is needed to adjust and tune these limits unless a fuzzy system, where the peak values of the fuzzy sets have the major effect, is developed first and its results are used. In the second case, we will have an additional increase in the software complexity of the classical controller. Although the development and tuning of the model to accommodate experts’ expectations might become easier to reach than in the first case, it is still complex compared to the fuzzy model. The ease by which the fuzzy models can be designed and tuned comes from the approximate nature of the models and available specialized software tools. 

Given the obtained results, the study confirms that the following objectives are met:
\begin{itemize}
	\item Reduce the time of the refrigerant charging procedure.
	\item Capture the intrinsic fuzziness of the process to reduce the controller complexity.
	\item Capture the decision-making process of a human operator.
	\item Obtain results that outperform classical crisp models.
	\item Perform parallel testing of many refrigerators. The system includes sensors, interfaces, hardware, and application software \deleted{graphical user interface} (\textit{GUI}).
	\item Automate the processes to save labor  
\end{itemize}

\subsection{Closely-Related Work}
\added{The attempt to identify related work succeeded in matching parts of the system proposed in this investigation. The identification comprises refrigerators testing procedures and automations~\cite{Adler06,kuijpers1988influence,radermacher1996domestic,djd00,bandarra2016energy,yang2015self,kocyigit2015fault,you2012optimizing,li2011fuzzy,csahin2012comparative,rashid2010design,pang2016strategy}, general-purpose automatic performance testing devices that can be used in refrigerators manufacturing, and automatic refrigerant charging devices~\cite{GTP,AGRAMKOW,VTECH18}. None of the identified systems attempt to fully-automate the refrigerant charging process of refrigerators.}

\added{In almost all investigations in~\cite{Adler06,kuijpers1988influence,radermacher1996domestic,djd00,bandarra2016energy,yang2015self,kocyigit2015fault,you2012optimizing,li2011fuzzy,csahin2012comparative,rashid2010design,pang2016strategy} (See Section~\ref{sec:2}), the purposes include optimizing the performance of refrigeration in terms of the lowest attained temperature, power consumption, optimal charge combinations and amounts, etc. In this paper, the proposed work is analyzed according to similar performance indicators, however, it differs in the approach and application. The proposed approach aims at mimicking the behavior of a human tester, during refrigerant charging, to reach the best possible performance of refrigeration. In addition, the application of the proposed approach is on performance testing of refrigerators during the manufacturing process with automated refrigarent amount increase.}

%\added{In~\cite{kuijpers1988influence}, Kuijpers et al. studies the influence of the refrigerant charge on the functioning of small refrigerating appliances. The authors in~\cite{bandarra2016energy} investigate achieving energy conservation in refrigeration systems by means of hybrid fuzzy adaptive control techniques. Yang et al. in~\cite{yang2015self} [8] propose a self-adjusting fuzzy logic controller for refrigeration systems. In~\cite{kocyigit2015fault}, the author develops fault and sensor error diagnostic strategies for a vapor compression refrigeration system by using fuzzy inference systems and artificial neural network. You et al. in~\cite{you2012optimizing}, propose a performance optimization approach to varying load of the refrigeration system based on fuzzy logic control. Pang et al. in~\cite{pang2016strategy} propose a strategy to optimize the charge amount of the mixed refrigerant for the Joule-Thomson cooler. Indeed, the purpose of the identified investigations is optimizing the performance of refrigeration in terms of the attained temperature, power consumption, optimal charge combinations and amounts, etc. In this paper, the proposed work is analyzed according to similar performance indicators, however, it differs in the approach and application. The proposed approach aims at mimicking the behavior of a human tester, during refrigerant charging, to reach the best possible performance of refrigeration. In addition, the application of the proposed approach is on performance testing of refrigerators during the manufacturing process.}

\added{Many general-purpose end-of-line Computerized Performance Tests (\textit{CPT}) devices, that can be used in refrigerators manufacturing, are available in the market. Example devices are the \textit{CPT} systems by \textit{AGRAMKOW}~\cite{AGRAMKOW}; the devices are equipped with advanced sensing, data acquisition, monitoring, calibration, analysis and other automation features. Supported automatic tests include temperature, pressure, power, air flow and humidity. Moreover, the \textit{CPT} devices can be used for short-term or long-term tests of refrigerators and other similar appliances. \textit{GALILIO} provides the \textit{CAPTOR K} data logger of field signals; the device is supported by \textit{RecData TJ} software package. The system is interfaced to Personal Computer; it performs performance tests on cooling units and other applications. The test is carried out by collecting the temperature, current curves and other signal of the units which have been tested using the \textit{Captor K} device. Test may last from a minimum of 1 Second up to maximum of 99 days~\cite{GTP}.}

\added{A variety of automatic refrigerant charging devices are available in the market~\cite{GTP,AGRAMKOW,VTECH18}. Such equipment can be classified per use, such as, commercial and domestic refrigeration. Devices vary in size, refrigerant type and charge weight. Moreover, such refrigerant charging devices are used in fast-paced assembly lines with limited process variability; they aim at providing automation, process data traceability and high charging accuracy. The main functions of such devices include refrigerant charging, leak testing, etc. 
}

\section{Conclusion}
\label{sec:10}

This study has introduced a new approach mainly based on fuzzy logic and inference for making the experimental human-performed control of the refrigerant-charging process of refrigerators intelligent and automated. The automation \added{is} \deleted{has been} done while accounting for the reduction of the time needed to finish the testing and performance evaluation. The approach is \deleted{has been} motivated by the importance of the charging process in terms of achieving optimal performance and \added{improved power consumption}. It has also been motivated by the natural fuzziness that arises in the basic description of the experimental testing procedure and by how human experts per-form the charging process. 

Based on the expert testing procedure (See Section~\ref{sec:4}); fuzzy and classical inference models \added{are} \deleted{have been} developed, \deleted{in this study and} \added{where} both \deleted{have been} show\deleted{n} the capability of representing the \deleted{behavior of an} intelligent human tester \added{behavior and} reflecting the viability of using inference methodology in \deleted{this} application. From the \deleted{general} perspective of better accommodating \added{for} \deleted{the} approximate \deleted{type of} human reasoning \added{in the application in hand} \deleted{and accounting for the application case under study}, the fuzzy inference models are preferable. \deleted{These} \added{the developed} models \added{are} \deleted{have been} shown to provide a \deleted{more} suitable representation of the natural fuzziness that arises in the problem statement, \deleted{and its described solution} and \deleted{the implementation of} the correlation principle implementation that governs the behavior of an intelligent human expert. 

Furthermore, the fuzzy model \deleted{has been} is shown (See Table~\ref{tbl:4}) to require less software, development and tuning complexity compared to the classical models. The approximate reasoning aspect and the availability of fuzzy inference tools permit the fuzzy system to be designed and tuned easily to accommodate any refrigerant charging situation and expert’s expectation. \added{Furthermore, inference tools} \deleted{It also} permit\deleted{s} the fuzzy models to be embedded in a computerized data acquisition\added{,} \deleted{;} monitoring \added{,} and decision-making system containing the necessary hardware and software parts with the needed interfaces.

The provided fuzzy models \added{are} \deleted{have been} built upon reasonable \deleted{yet general} assumptions regarding the trend of temperature change versus time\added{,} and the speed of data acquisition. Given a specific refrigerant charging situation\added{,} and \added{the} corresponding temperature versus time curve, then the data acquisition speed could be adjusted in a manner to make the developed fuzzy model match that situation. \deleted{Whether this cannot be done, then these} \added{Indeed,} the proposed models can be tuned to provide a better match. The tuning can be applied to the ranges of the used variables, the \textit{THEN} parts of the rules\added{,} and the location of the supports and peak values of the fuzzy sets. This kind of tuning causes a change in the time needed to stop the testing for every value or range of values of $Dtemp$ (See Section~\ref{sec:7}). \added{With no doubts} \deleted{But of course}, the principles explained in Section~\ref{sec:5} need always \added{to} be observed. Hence, the provided solution procedures are general and remain valid.   

The obtained results confirmed that the fuzzy controller, and in its best case, can achieve a test time saving of 70.6\%. \deleted{This} \added{The saving} leads to a constant energy consumption of 1 kWh per tested unit. The best \added{achieved} test\added{-}time saving results\added{,} \deleted{achieved} using the fuzzy controller\added{,} outperform\added{s} those of the crisp controllers. However, the best-case crisp and worst-case fuzzy controller perform in an almost identical manner.

Future work \deleted{should} deal\added{s} with the application of the introduced approach in various refrigerant-charging situations\added{,} and with the tuning and improvement of the provided fuzzy models to try to make them competitive with the performance of well-trained human experts. Although the minimization of the time factor is important, this minimization should not be done at the expense of detecting a temperature that significantly differs from the smallest possible one for every amount of refrigerant. If \deleted{this} \added{a value around the smallest possible temperature} is not reached, then the \added{calculated \textit{COP}} \deleted{coefficient of performance  calculated} for every specific refrigerant amount would turn out to be wrong\added{, and leads} \deleted{leading} to a wrong decision on the amount of needed refrigerant. Hence, the specific charging situation, the balancing between the time reduction and accuracy of the detected temperature, as well as the experts’ opinion and expectation, need to be given careful considerations to obtain reliable and efficient fuzzy inference model. \added{In addition, Future work can include investigating improvements towards fault-prone manufacturing using model-reduction techniques~\cite{YJHM2014,WEI2014316}}.

\begin{table*} 
	%\begin{center}
		\caption{Comparison of crisp and fuzzy controllers.} 
		\label{tbl:4}
		\small
		\begin{tabular}{c|c|c}	
			\hline 
			
			\textbf{Indicator} & Crisp Controller & Fuzzy Controller\\
			
			& Equation~\ref{eqn:7} & Table~\ref{tbl:1}\\
			\hline
			
			Procedure required to complement the Software tool & Complex & Simple\\
			\hline
			
			Development and tuning under tools such as LabVIEW& Complex & Simple\\	
			\hline
			
		\end{tabular}
		
	%\end{center}	
\end{table*}

\section*{Acknowledgments}
The authors would like to thank Mr. M. El-Khalili, a Senior Mechanical Engineer and Market Area Director at Eberspächer Sütrak GmbH, Germany, for the help and advice he provided in some paper related issues. \added{In addition, the authors are grateful for the thorough reviews, by the editor and the anonymous reviewers, that enabled great improvements to this paper.}

%\bibliographystyle{unsrt}
%\bibliography{GCTref}   % name your BibTeX data base

\bibliographystyle{unsrt}
\bibliography{GCTref}   % name your BibTeX data base

%\begin{thebibliography}{}
% etc
%\end{thebibliography}

%\listofchanges[style=<list|summary>]

\section*{Appendix. List of Acronyms and Symbols}

\begin{table} [htb]
	
	%\caption{\added{List of Acronyms and Symbols and their definitions}}
	\label{tbl:1}
	
	%\begin{center}
	\small
	
	\begin{tabular}{l | l}	
		\hline 
		\textbf{Acronym or Symbol} & \textbf{Definition} \\
		\hline
		\hline
		
		{\textcelsius} & Degree Celsius \\
		\hline
	
		COI & Compositional Rule of Inference \\
		\hline
		
		COP & Coefficiant of Performance \\
		\hline
		
		CPT & Computerized Performance Test\\
		\hline
		
		DAQ & Data Acquisition \\
		\hline
		
		DTemp & Temperature Change Variable \\
		\hline
		
		E & Energy\\
		\hline
		
		GUI & Graphical User Interface\\
		\hline

		kWh & Kilo Watts per Hour\\
		\hline

		max & Maximum\\
		\hline
		
		$\wedge$ & Minimum Operator\\
		\hline
		
		MF & Membership Function\\
		\hline
		
		MISO & Multiple Input Single Output\\
		\hline
		
		NI & National Instruments\\
		\hline

		P & Power\\
		\hline

		PR & Performance\\
		\hline
		
		S4 & Suction Tube \\
		\hline

		t & time\\
		\hline

		TH & Throughput\\
		\hline
		
		TpD & Tests per 12-hour work Day\\
		\hline

		TpH & Tests per Hour\\
		\hline
		
		TS & Time Saving\\
		\hline
		
		TT & Test Time \\
		\hline
		
		Time & Observation Time Variable \\
		\hline

	\end{tabular}
	%\end{center}
	
\end{table}

\parpic{\includegraphics[width=1.1in,,keepaspectratio]{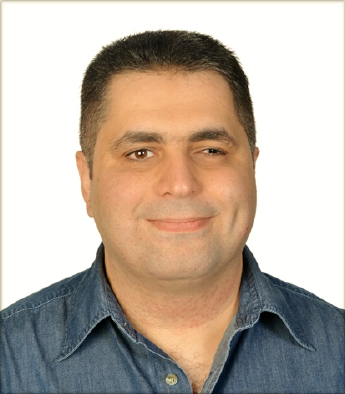}}
\noindent {\bf Issam W. Damaj} is an Associate Professor of Computer Engineering at the Electrical and Computer Engineering (ECE) Department at American University of Kuwait (AUK). Since 2011, he has been the ABET Accreditation Lead and Institutional Representative of the ECE Department. He was the Founding Chairperson of the ECE Department and Program Lead between September 2009 and August 2016.  He was awarded a Doctor of Philosophy Degree in Computer Science from London South Bank University, London, United Kingdom in 2004. He received a master’s Degree in Computer and Communications Engineering, in 2001, from the American University of Beirut. He received a bachelor’s Degree in computer engineering from Beirut Arab University in 1999. His research interests include hardware/software co-design, embedded system design, automation, Internet-of-things, fuzzy systems, formal methods, and engineering education. He is a Program Evaluator (PEV) with ABET Engineering Accreditation Commission (EAC), a Senior Member of the IEEE, a Professional Member of the ASEE, and a member of the Order of Engineers in Beirut. He maintains an academic website at www.idamaj.net.

\parpic{\includegraphics[width=1in,clip,keepaspectratio]{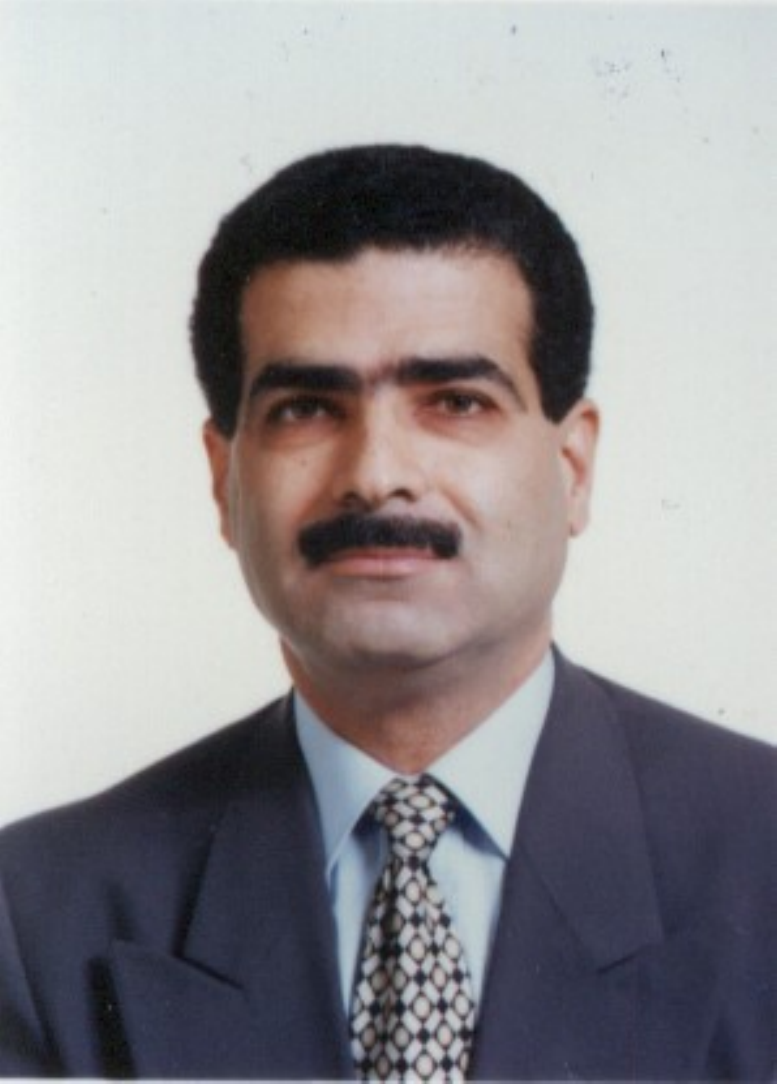}}
\noindent {\bf Jean J. Saade} Jean J. Saade was born in Lebanon in 1954. He received the B.Sc. degree in Physics from the Lebanese University, Lebanon, in 1979 with a scholarship to pursue Ph.D. studies. He then received the M.Sc. degree in Electrical Engineering in 1982 from Boston University, MA, USA, and a Ph.D. degree, also in Electrical Engineering, from Syracuse University, NY, USA, in 1987. He taught for four years at Syracuse University while being a Ph.D. candidate and served for one year as a Visiting Assistant Professor at the same university after graduation. In 1988, he joined the American University of Beirut, Lebanon, and is currently a Professor at the Department of Electrical and Computer Engineering, teaching communication, signal processing and fuzzy logic courses. Dr. Saade published a good number of papers in international journals and conferences addressing the mathematical aspects and application of fuzzy sets and logic in signal detection, radar and robot navigation. He is a member of the European Society for Fuzzy Logic and Technology (EUSFLAT).

\parpic{\includegraphics[width=1in,clip,keepaspectratio]{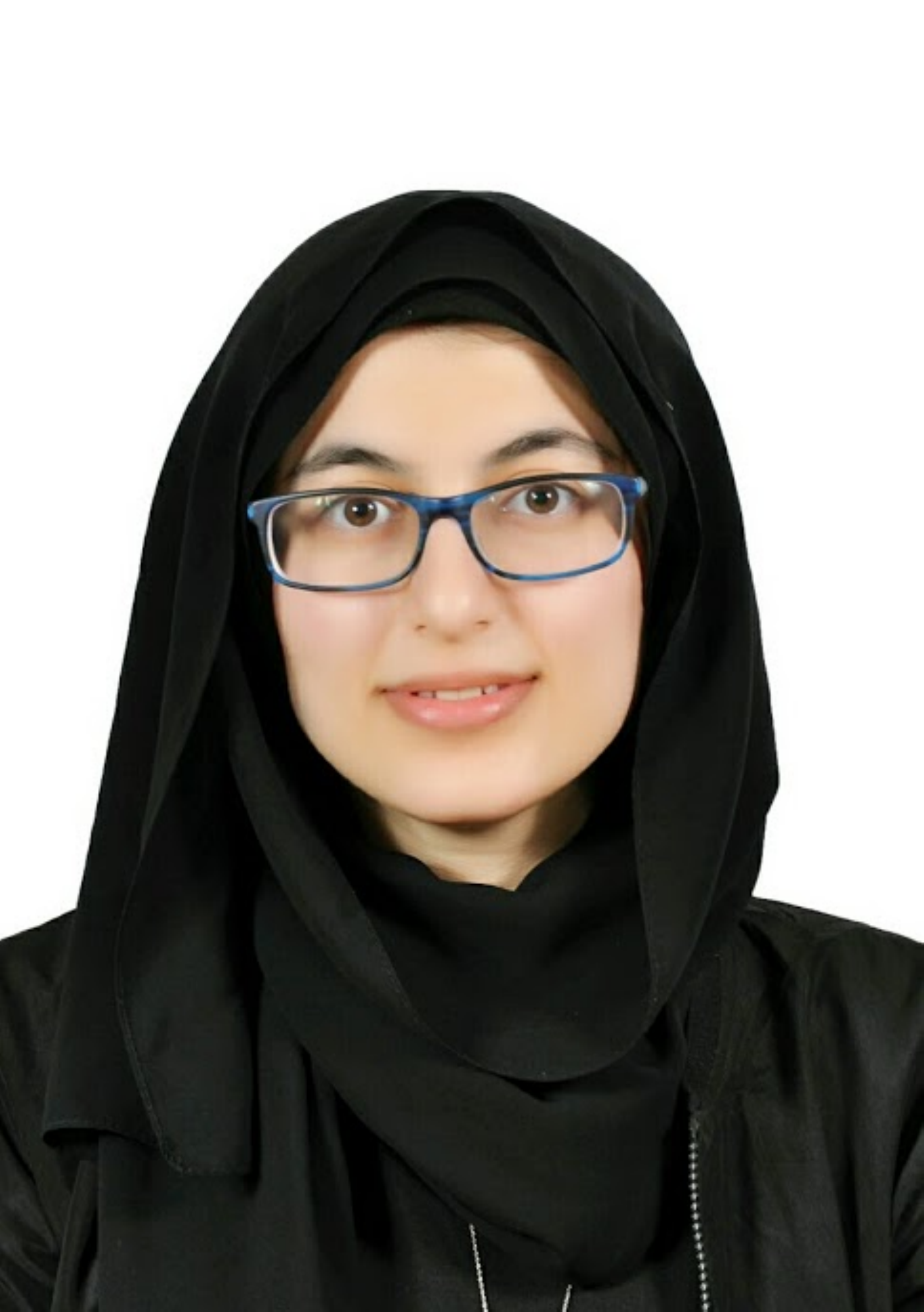}}
\noindent {\bf Hala R. Al-Faisal} received the bachelor’s degree in computer engineering from the American University of Kuwait in 2016, and is currently pursuing master’s degree at Kuwait University in the same field. Current research interests include artificial intelligence, and intelligent control systems.\\
\\

\parpic{\includegraphics[width=1.1in,clip,keepaspectratio]{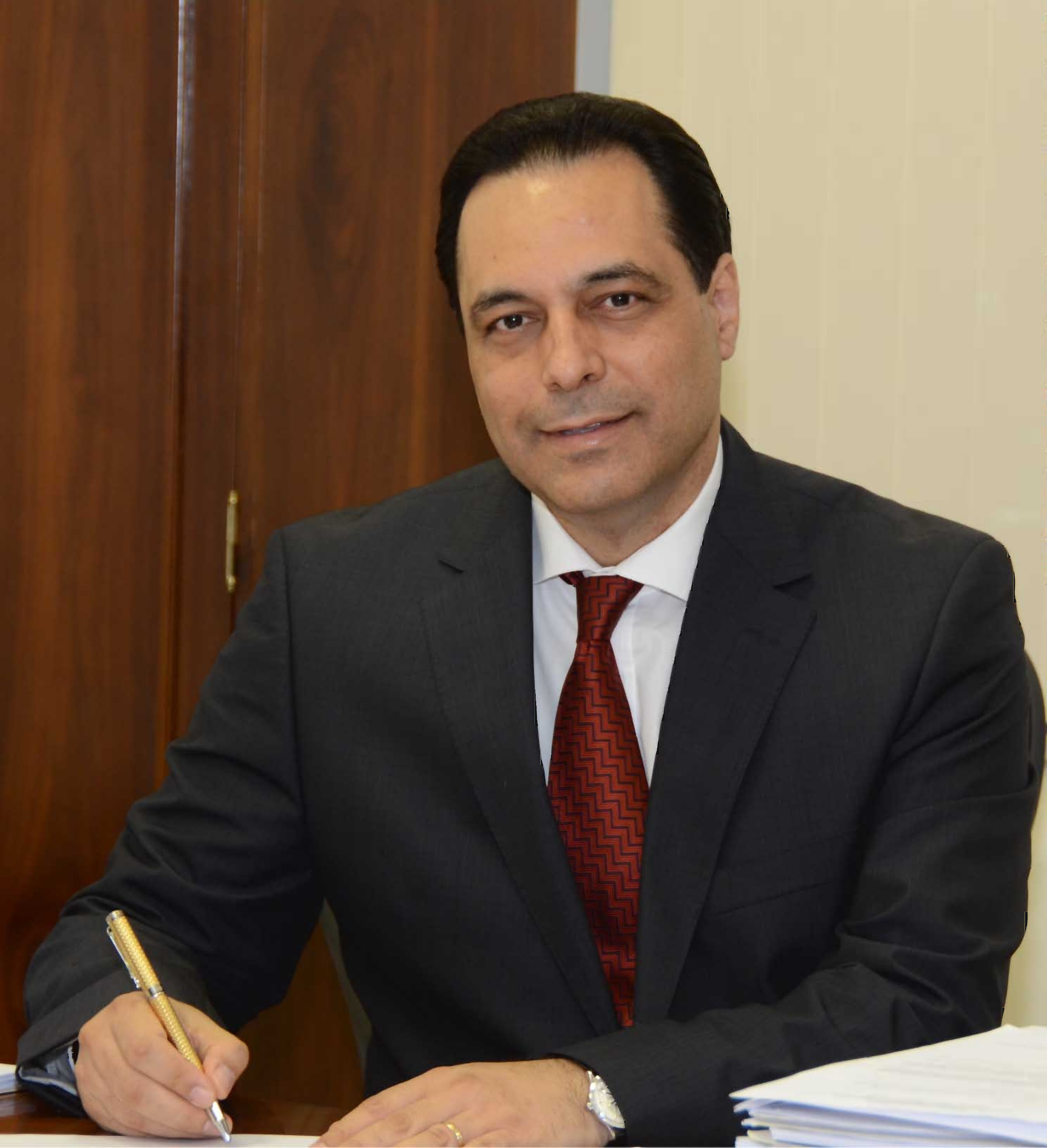}}
\noindent {\bf Hassan B. Diab} received his B.Sc. (with Honors) in Communications, M.Sc. (with Distinction) in Systems Engineering, and Ph.D. in Computer Engineering. He joined the American University of Beirut (AUB) in 1985 and is a Professor of Electrical and Computer Engineering at the Faculty of Engineering and Architecture. He has over 140 publications in internationally refereed journals and conferences. His research interests include cryptography on high performance computer systems, modeling and simulation of parallel processing systems, embedded systems, reconfigurable computing, as well as several areas in education. He has supervised/co-supervised over 80 research projects, including 30 Masters Theses. He served as Associate Editor or member of Advisory/Editorial Board on five international journals. Professor Diab received over 20 international and regional awards. He has served as Founding Dean of the College of Engineering and Founding President during 2004-2006 at Dhofar University. Effective October 2006, Professor Diab was appointed as Vice President for Regional External Programs (REP) at AUB. He is a Founding Member of the first Arab Computer Society established in 2001 as well as the Founding Member of the IEEE Student Branch at AUB in 1997. Professor Diab is a Fellow in the IEE and the IEAust, as well as a Senior Member of the IEEE. During June 2011 – February 2014, he was appointed as Minister of Education and Higher Education in the Lebanese Cabinet to take over the largest ministry in Lebanon. Since July 2013, he returned to his position as VP at AUB.

\end{document}